\documentclass[iop,revtex4]{emulateapj}
\usepackage{lscape}

\usepackage{color}
\usepackage{ulem}

\shorttitle{On the nature of MAXI~J1305$-$704}
\shortauthors{Shidatsu et al.}

\begin{document}

\title{Accretion disk and ionized absorber of the 9.7-hour dipping black hole binary MAXI~J1305$-$704}

\author{M. Shidatsu\altaffilmark{1}, Y. Ueda\altaffilmark{1}, S. Nakahira\altaffilmark{2}, 
C. Done\altaffilmark{3}, K. Morihana\altaffilmark{4}, M. Sugizaki\altaffilmark{5}, 
T. Mihara\altaffilmark{5}, T. Hori\altaffilmark{1}, H. Negoro\altaffilmark{6}, 
N. Kawai\altaffilmark{7}, K. Yamaoka\altaffilmark{8}, 
K. Ebisawa\altaffilmark{9}, M. Matsuoka\altaffilmark{2}, M. Serino\altaffilmark{5}, 
T. Yoshikawa\altaffilmark{1}, T. Nagayama\altaffilmark{10}, N. Matsunaga\altaffilmark{11}}
\email{shidatsu@kusastro.kyoto-u.ac.jp}

\altaffiltext{1}{Department of Astronomy, Kyoto University, Kitashirakawa-Oiwake-cho, 
Sakyo-ku, Kyoto 606-8502, Japan}
\altaffiltext{2}{ISS Science Project Office, Institute of Space and Astronautical Science (ISAS), 
Japan Aerospace Exploration Agency (JAXA), 2-1-1 Sengen, Tsukuba, Ibaraki 305-8505, Japan}
\altaffiltext{3}{Department of Physics, University of Durham, South Road, Durham, DH1 3LE, UK}
\altaffiltext{4}{Nishi-Harima Astronomical Observatory, Sayo-cho, Hyogo 6
79-5313, Japan}
\altaffiltext{5}{MAXI team, Institute of Physical and Chemical Research (RIKEN), 
2-1 Hirosawa, Wako, Saitama 351-0198, Japan}
\altaffiltext{6}{Department of Physics, Nihon University, 1-8-14 Kanda-Surugadai, Chiyoda-ku, 
Tokyo 101-8308, Japan}
\altaffiltext{7}{Department of Physics, Tokyo Institute of Technology, 2-12-1 Ookayama, 
Meguro-ku, Tokyo 152-8551, Japan}
\altaffiltext{8}{Solar-Terrestrial Environment Laboratory, Nagoya University, Furo-cho, 
Chikusa-ku, Nagoya, Aichi 464-8601, Japan}
\altaffiltext{9}{Institute of Space and Astronautical Science (ISAS), 
Japan Aerospace Exploration Agency (JAXA), 3-1-1 Yoshino-dai, Chuo-ku, Sagamihara, 
Kanagawa 252-5210, Japan}
\altaffiltext{10}{Department of Astrophysics, Nagoya University, Furo-cho, Chikusa-ku, Nagoya 464-8602, Japan}
\altaffiltext{11}{Department of Astronomy, School of Science, The University of Tokyo, 7-3-1 Hongo, Bunkyo-ku, 
Tokyo 113-0033, Japan}

\begin{abstract}
We report the results from X-ray studies of the newly discovered black
hole candidate MAXI J1305-704 based on {\it Suzaku} and {\it Swift}
observations in the low/hard and high/soft states, respectively. The
long {\it Suzaku} observation shows two types of clear absorption
dips, both of which recur on a dip interval of $9.74\pm 0.04$ hours,
which we identify with the orbital period. There is also partially
ionized absorption in the non-dip (persistent) emission in both the 
high/soft state and, very unusually, the low/hard state. However, this
absorption (in both states) has substantially lower ionization than
that seen in other high inclination systems, where the material forms
a homogeneous disk wind. Here instead the absorption is most probably
associated with clumpy, compact structures associated with the dipping
material, which we see uniquely in this source likely because we view it 
at a very large inclination angle. A large inclination angle is also favored, 
together with a low black hole mass, to explain the high disk temperature 
seen in the fairly low luminosity high/soft state, as Doppler boosting 
enhances the disk temperature at high inclination. The disk radius 
inferred from these data is significantly smaller than 
that of the soft component seen in the low/hard state, supporting models 
where the disk is truncated at low luminosities. 
We find, however, that the lack of variability power on time scales of 
$\sim 50$~sec in the {\it Suzaku} low/hard state data is difficult to explain, 
even with a low mass black hole.

\end{abstract}


\section{Introduction}

Transient black hole X-ray binaries (BHXBs) are the best laboratories
to study the physics of the accretion flow in a wide range of the mass
accretion rate.  They make drastic changes in spectral properties
during their outburst while exhibiting orders of magnitude increase and
decrease of their X-ray luminosity, suggesting that the geometry of
inner disk differs significantly according to the mass accretion rate
\citep[e.g.,][and references therein]{mcc06,don07}.  In low X-ray
luminosity phases, they show a relatively hard, power-law shaped
spectrum with a photon index of less than 2.0 and an exponential
cutoff at $\approx$100 keV, which is generally interpreted as the disk
emission Comptonized by thermal electrons in the surrounding
corona. This state is called ``low/hard state'', in which the inner 
part of standard disk \citep{sha73} is thought to be truncated. Following 
the rapid increase of the X-ray luminosity in outbursts, they undergo a 
state transition to the so-called ``high/soft state'' typically at a mass
accretion rate of $\sim 0.1 L_{\rm Edd}$ ($L_{\rm Edd}$ is the Eddington 
Luminosity: $4\pi G m_{\rm P} c M_{\rm BH}/\sigma_{\rm T}$, where $G$, 
$m_{\rm P}$, $c$, $M_{\rm BH}$, and $\sigma_{\rm T}$ represent the 
gravitational constant, proton mass, speed of light, black hole mass, 
and Thomson scattering cross-section, 
respectively), in which the soft X-ray flux is dominated by thermal 
emission from the standard accretion disk.  Many previous studies showed 
that the inner disk radius stays constant during the high/soft state
\citep[see e.g.,][]{ebi93}. This suggests that the disk extends down
to the inner most stable circular orbit (hereafter ISCO) in the
high/soft state.

The properties of fast time variabilities are also remarkably
different between the low/hard state and the high/soft state. In the
low/hard state, BHXBs show noisy light curves on time scales of up to
hundred seconds. Their power density spectra (PDSs) are roughly
characterized with the so-called band limited noise with flat profile
in the $\nu P_{\nu}$ spectrum ($\nu P_{\nu} \propto \nu^0$) within the 
low and high frequency break, below and above which the power declines 
as $\nu P_{\nu} \propto \nu^1$ and $\nu P_{\nu} \propto \nu^{-1}$, 
respectively.  This noise
profile is better described as a superposition of multiple Lorentzians
\citep{bel90, now00, bel02}. \citet{neg01} reported that these 
structures are reproduced by the superposition of ``shots'' (flare like 
events) seen in the low/hard state light curve, which are thought to be 
related to density fluctuation of advection-dominated accretion flow 
inside the inner edge of the standard disk \citep[e.g.,][]{man96}.
In contrast, rapid variability is 
typically weak in the high/soft state, where the constant standard
disk emission dominates the X-ray flux \citep[e.g.,][]{hom01}. The
low-frequency break of the band-limited noise seen in the low/hard
state moves toward higher frequencies as the X-ray luminosity 
increases, and the profile of PDS is smoothly connected to those in
the high/soft state through the intermediate or very high
state \citep{van04, axe05}. These characteristics support the idea
that the standard disk is truncated and the inner edge moves inward to
reach ISCO as the luminosity increases \citep{ing12}.

BHXBs with a relatively high inclination angle are particularly
interesting objects because they give us key information to uncover
the structure of the outer accretion disk.
In the high/soft state, these sources often exhibit highly ionized
blue-shifted absorption lines that originate in the ``disk wind''
outflowing from the outer region of the accretion disk
\citep[e.g.,][]{ued98, kot00, mil06a, kub07, pon12}. 
Importantly, the mass loss rate of a disk wind is comparable
with or even several to a few dozens times larger than the mass accretion
rate \citep{ued04, nei11}, which suggests that the disk wind would
also affect the properties of inner region of the disk and play a
critical role on accretion disk physics.

High inclination X-ray binary systems often show quasi-periodic
dips accompanied by spectral hardening in their light curves. It is
generally believed that the dips are caused by absorption of the X-ray
emission from the central source with dense structure in outer disks
such as the ``bulge'', which is formed by the accretion stream from
the companion star impacting the rim of the disk \citep[see
e.g.,][]{whi82}. Previous studies showed that dipping spectra are well
reproduced by a partial absorption by neutral material
\citep[e.g.,][]{mar93}. While this approach was successfully
applied in many sources including both neutron star and black hole
X-ray binaries, another explanation has recently been proposed;
\citet{boi05} and \citet{dia06} successfully described both
non-dipping and dipping spectra of neutron-star low mass X-ray
binaries, using a single photo-ionized absorption model with different
column densities and ionization parameters. It is suggested that
absorption dips are generally caused by ionized gas of lower
ionization state and higher column density than the disk winds (see
section~5.2).

MAXI~J1305$-$704 is an X-ray transient discovered on 2012 April 9 
\citep{sat12} with {\it MAXI}/GSC 
\citep[Monitor of All-sky X-ray Image/Gas Slit Camera;][]{mat09}. 
The monitoring results with the GSC
suggest that the source is likely a black hole X-ray binary, as its
hardness-intensity diagram showed a q-shaped hysteresis over the whole
outburst and the spectrum during the soft phase is well modelled with
thermal emission from the standard disk like those of typical
BHXBs in the high/soft state \citep{mor13}. A lot of follow-up
observations were triggered in X-ray and other wavelengths. 
Multiple {\it Swift} X-ray Telescope (XRT) observations discovered dips in 
the X-ray light curves, whose interval has been however still
controversial; 1.5 hours and 2.7 hours were suggested by \citet{ken12}
and \citet{kuu12}. {\it Swift}/XRT also detected strong ionized absorption
lines, likely originated in the disk wind \citep{mil12a}. {\it Chandra}
HETGS discovered complex absorption feature around 1 keV, which can be
reproduced by ionized iron-L absorption lines \citep{mil12b}.  Those
dips and absorption profiles strongly indicate that the source has a
large inclination angle, although the precise value is not determined
yet. The optical and near infrared counterparts were also detected in
the observations performed about a few days after the start of the
outburst \citep{gre12}.

In this paper, we present the results of a {\it Suzaku} TOO (Target of
Opportunity) observation of MAXI~J1305$-$704 performed during the
low/hard state to investigate the detailed properties of
the accretion flow, dips, and ionized absorbers in a low mass
accretion rate. The data obtained from a {\it Swift}/XRT observation 
during the high/soft state are also analyzed to be compared with
the {\it Suzaku} results.
In addition, we report the near infrared observations with the 1.4m
telescope of Infrared Survey Facility ({\it IRSF}) performed
quasi-simultaneously with the {\it Suzaku} observation and in an earlier
epoch when MAXI~J1305$-$704 was in the high/soft state. Errors
represent the 90\% confidence range for a single parameter in the
following sections. We refer to the table by \citet{and89} as the
solar abundances throughout the paper. 

\section{X-ray Observation and Data Reduction}

\subsection{{\it Suzaku} Observation in the Low/hard State}

We observed MAXI J1305$-$704 with {\it Suzaku} \citep{mit07} from 2012 July
20 18:10:29 (UT) to 22 00:30:23 for a net exposure of $\approx$40
ksec. This was carried out as a TOO observation based on the
monitoring by {\it MAXI}/GSC. {\it Suzaku} carries X-ray CCD camera called the 
X-ray Imaging Spectrometer (XIS), operated in the energy range of 
0.2--12 keV, and a non-imaging collimated instrument called the 
Hard X-ray Detector (HXD), which consists of PIN silicon diodes 
sensitive to 10--70 keV and gadolinium silicon oxide (GSO) crystal 
scintillators covering 40--600 keV. The XIS consists of two 
frontside-illuminated (FI) chips (XIS-0 and XIS-3) and a 
backside-illuminated (BI) chip (XIS-1), which has a larger effective 
area than the FI chips below $\approx$ 1.5 keV and a higher 
sensitivity to low energy X-rays. In this observation, the 1/4 window 
option was employed for the XIS. The actual observed count rate was 
$\approx$ 5 counts sec$^{-1}$ on average, which is low enough that 
we can ignore any effects by pileup. The {\it Suzaku} observation 
(MJD 56128--55130) corresponds to the period after the spectral 
hardening at the end of the outburst in 2012 June \citep{mor13}, 
suggesting that the source was in the low/hard state in our 
observation.

\setcounter{footnote}{0}
We utilized the cleaned event data produced by the pipeline
processing version 2.7.16.33, and reduced them with HEASOFT version 6.12 and
Calibration Database (CALDB) released on 2012 October 5. The source
events of the XIS were extracted from a circular region centered on the
source position with a radius of 1.9'. The background was taken from a
circular region with the same radius in a source-free area. For the non
X-ray background of the HXD, we used the modelled background files
provided by the {\it Suzaku} team. The modelled spectrum of the cosmic X-ray
background was subtracted from the PIN data, but not from the GSO data,
because its contribution is less than 0.1\% of the total background rate
of the
GSO\footnote{http://www.astro.isas.ac.jp/suzaku/analysis/hxd/gsonxb/}.
The PIN and GSO data were corrected for the dead time with {\tt
hxddtcor}. The XIS response matrix and ancillary response files were
created with the {\tt xisrmfgen} and {\tt xissimarfgen}, respectively,
to be used in our spectral analysis. We utilized {\tt
ae\_hxd\_pinxinome11\_20110601.rsp} as the response file of PIN, and
{\tt ae\_hxd\_gsoxinom\_20100524.rsp} and {\tt
ae\_hxd\_gsoxinom\_crab\_20100526.arf}
\footnote{http://www.astro.isas.ac.jp/suzaku/analysis/hxd/gsoarf2/} as
those of GSO. We combined the spectra and response files of the
FI-XISs (XIS-0 and XIS-3) to improve statistics. The data in the
1.7--1.9 keV band were always ignored in the spectral fits due to the
systematic uncertainties in the instrumental Si-K edge. A 1\% systematic
error was included in each bin of the XIS and HXD spectra to account for
possible calibration uncertainties.

The spectra of the FI-XISs, BI-XIS, and HXD were simultaneously
fitted in the spectral analysis. The cross-normalization 
of the HXD with respect to the FI-XISs was fixed at 
1.16\footnote{http://www.astro.isas.ac.jp/suzaku/doc/suzakumemo/suzaku\\memo-2008-06.pdf}.
We corrected for cross-calibration errors in the energy 
responses between the FI-XISs and BI-XIS, as we found that our 
FI-XIS data resulted in significantly harder spectra than 
the BI-XIS ones from the individual spectral analysis, probably 
due to uncertainties in modelling the contamination on the 
XIS window filters.
To examine trends of these uncertainties in the period near our 
observation, we analyzed two {\it Suzaku} archival data of the 
blazar PKS 2155$-$304 observed on 2012 April 27-29 and October 30-31, 
both of which were operated with the same (1/4) window option. 
We created time-averaged FI-XIS and BI-XIS spectra separately 
for the two epochs, using the same versions of HEASOFT and CALDB as 
those applied in the analysis of the MAXI J1305$-$704 data. 
The spectra were fitted with an absorbed power-law model, in which 
the photon indices were linked between the FI-XIS and BI-XIS data.
We found that the FI-XIS spectra show larger $N_{\rm H}$ than the
BI-XIS ones by $\Delta N_{\rm H} = 1.2 \times 10^{20}$ cm$^{-2}$ and 
$\Delta N_{\rm H} = 2.0 \times 10^{20}$ cm$^{-2}$ in the April and 
October observations, respectively. Similarly, for the case of 
MAXI J1305--704, we estimated difference of 
$\Delta N_{\rm H} = 3 \times 10^{20}$ cm$^{-2}$. To account for this offset, 
we unlinked the column density of the neutral absorption along with 
the flux normalization between the FI-XIS and BI-XIS spectra in the 
simultaneous fit.
In the following section we show the column density obtained from 
the BI-XIS spectrum as the best estimated value. 
The inclusion of this correction is found to significantly improve 
the quality of the fit, although it does not affect the conclusion 
of this paper.

MAXI J1305$-$704 is located near a bright source 4U 1254$-$690 with a
separation angle of 1.41$^\circ$ and the GSO flux can be contaminated
by the emission from this nearby source \citep{tak07}.  However, we
confirmed that the contamination is completely negligible as the
source is more than several orders of magnitude fainter than our
target above 50 keV by considering the previous spectral study of
4U 1254$-$690 \citep{dia09}.

\subsection{{\it Swift}/XRT Observations in the High/soft State}

Since its discovery, MAXI~J1305$-$704 was observed with {\it Swift}/XRT many
times. In order to compare the {\it Suzaku} data in the low/hard 
state with {\it Swift} ones in the high/soft state, we analyzed the 
data of {\it Swift}/XRT obtained from 2012 April 19 13:19:53 to
21 17:03:00 (UT). This is one of the longest ($\approx$ 10 ksec)
{\it Swift}/XRT observations for this source in the high/soft state. Using
this dataset, \citet{mil12a} reported the existence of a strong iron-K
absorption line around 6.6 keV. In this observation, XRT was operated
with the 1-dimensional Window Timing mode. The data are not affected
by pile-up, because the averaged count rate ($\approx 30$ counts
sec$^{-1}$) is much lower than the maximum pileup-free count rate
\citep[100 counts sec$^{-1}$;][]{rom06}.

We used {\it Swift}/XRT archival data and performed the standard reduction
with {\tt xrtpipeline}. The source events were extracted from a box
region of 40 pixels $\times$ 30 pixels along with the X- and Y-axis in
the detector coordinates, respectively, with the center located at the
target position. The background region was defined as two boxes of 40
pixels $\times$ 30 pixels in the source-free area at the same distance
from the target position. We included 3\% systematic error in each
spectral bin to absorb possible calibration 
uncertainty\footnote{http://heasarc.gsfc.nasa.gov/docs/heasarc/caldb/swift/docs/\\xrt/SWIFT-XRT-CALDB-09\_v16.pdf}. 
We utilized a response matrix file, {\tt swxwt0to2s6\_20010101v014.rmf}, taken from
the {\it Swift} CALDB provided on 2012 October 5. The ancillary response
file is created by using {\tt xrtmkarf} with the exposure file
produced in the pipeline tool.

\begin{figure}
\epsscale{1.05}
\plotone{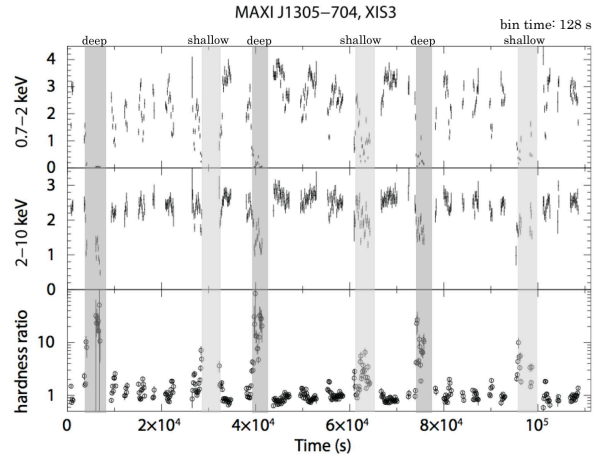}
\caption{The XIS-3 light curves in 0.7--2 keV and 2--10 keV, and their ratio 
in 128 sec binning, from the top to bottom. The shadowed regions represent 
the periods of the deep dips (dark gray) and the shallow dips (light gray). 
\label{fig_lcsuzaku}}
\end{figure}

\section{Analysis and Results}

\subsection{{\it Suzaku} light curve and dip feature}

\begin{figure}
\epsscale{0.95}
\plotone{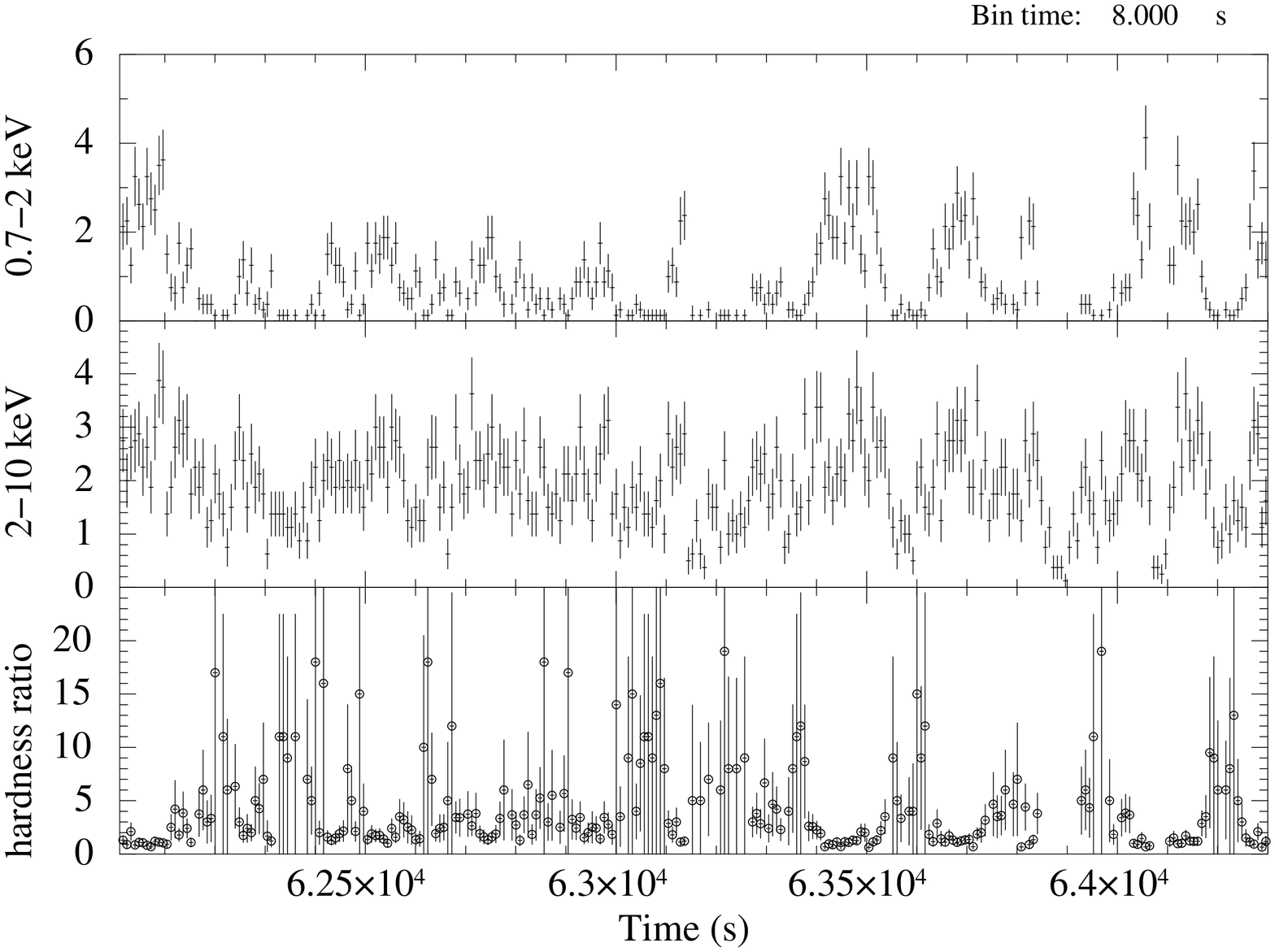}
\epsscale{1.0}
\plotone{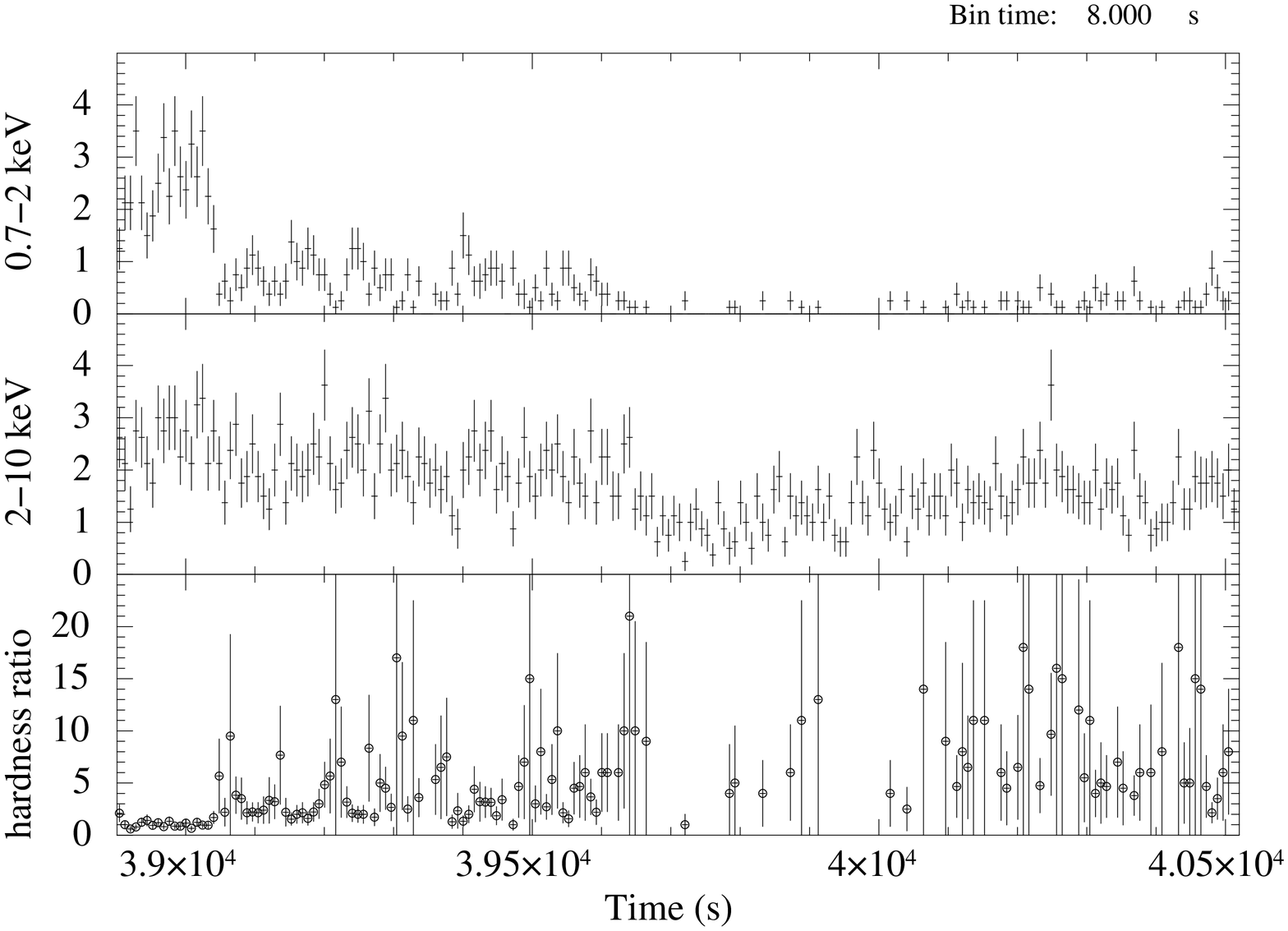}
\caption{Same as Fig.~\ref{fig_lcsuzaku} in 8 second bins. 
The upper and lower panels present the phases of a shallow dip 
and a deep dip, respectively. \label{fig_lcshort}}
\end{figure}

Figure~\ref{fig_lcsuzaku} shows the {\it Suzaku} XIS-3 light curves in the
soft (0.7--2 keV) and hard (2--10 keV) bands together with their
hardness ratio in 128 sec binning. The light curve is highly
variable particularly in the soft band, suggesting that the
variability is mainly caused by absorption. We can see two dipping features 
with different mean hardness ratios ($\approx$5--10 for the softer ones 
and $\approx$20--30 for the harder ones), in which more than 80\% of the 
averaged flux is lost in the soft band, and those dips with similar mean 
hardness ratios are observed almost periodically.
Here we define the start and end times of the dips as the points 
at which the hardness ratio crosses a value to 2.6 upward and downward 
in Fig.~\ref{fig_lcsuzaku}. We then call the dips whose peak hardness ratios 
reach 20 in 128 sec binning as ``deep dips'' and the other softer ones as 
``shallow dips''. In shorter time scales, the shallow dips have small 
variabilities with typical time scale of a few minutes, while this 
behavior is not significant in the deep dips (Figure~\ref{fig_lcshort}).

We find each dip recurrently occurs with a period of 9.74 $\pm$ 0.04
hours, which is calculated from the intervals of the start times of
deep dips obtained in the XIS-3 hardness ratio with 64 sec bin. 
The error is estimated by propagating the uncertainty of each measured 
start time, which is assumed as half width of time bins in the light 
curve (32 sec). This interval, instead of 1.5 or 2.7 hours suggested 
by {\it Swift}/XRT observations with shorter exposure \citep{ken12,kuu12}, 
likely corresponds to the orbital period. The light curves have data 
gaps with durations of 0.8--1.1 hours, which are not exactly periodic. 
The durations of the deep and shallow dips (1.0--1.7 hours) are comparable 
or larger than those of the data gaps. Because the hardness ratio never 
exceeds 2.6 outside of the dip phases we identified, it is unlikely that 
the actual period is shorter than 9.74 hours and we miss other dip events 
in these data gaps. We also note that dips do not always appear precisely 
in the same orbital phase, and the interval between the first and second 
deep dips actually 50 sec longer than that of second and third deep dips, 
when we measure them from the light curve with shorter time bins. 
The interval from a deep dip to the next shallow dip is derived to be 
$6.38 \pm 0.04$ hours, which is obtained by averaging the intervals between 
the beginning of deep dips and those of the following shallow dips. 
The durations of the deep and shallow dips, however, are not precisely 
constant, and consequently their intervals are slightly different event 
by event.

\begin{figure}
\plotone{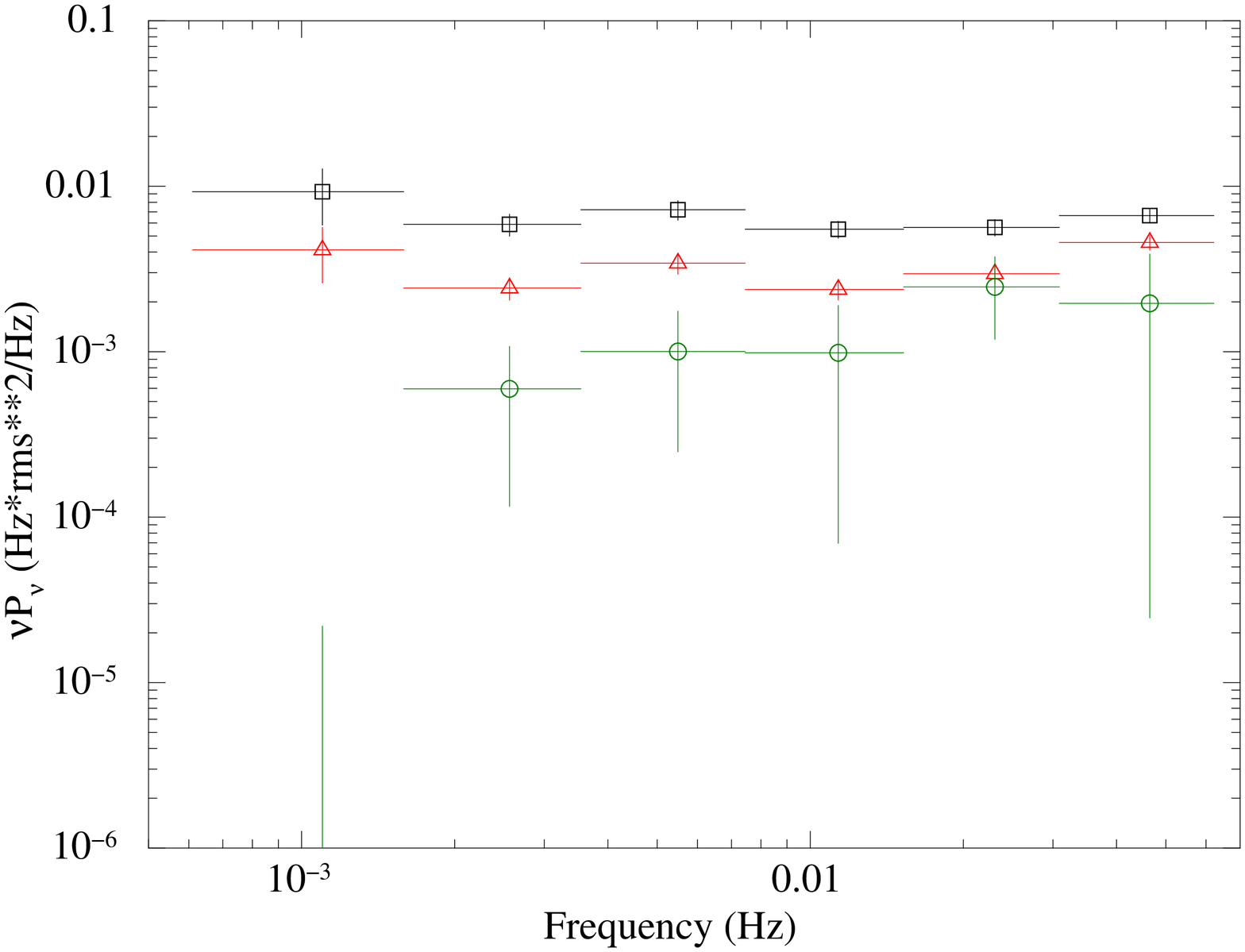}
\caption{The XIS power density spectra in the 0.7--1 keV (black, square), 
1--5 keV (red, triangle), and 5--10 keV (green, circle) band, created 
by using the XIS-0$+$XIS-1$+$XIS-3 light curve in the non-dip phases 
with 8 sec bins. They are normalized in the way that their integral gives
the squared root mean squared fractional variability. White noise is 
subtracted.
\label{fig_powspec}}
\end{figure}

Figure~\ref{fig_powspec} shows the normalized power density spectra 
in the non-dip 
phases in the 0.7--1 keV, 1--5 keV, and 5--10 keV bands, calculated 
from the combined light curve of all three XISs with 8 sec bins.  We
find that the softer energy band has much larger power than those of
the harder bands in the frequency range of $10^{-3}$ to $5 \times
10^{-2}$ Hz. This suggests that the power is likely dominated by the
residual variability of absorption that exists even outside the dips.
The source has only small intrinsic power ($\approx 1 \times 10^{-3}$
rms$^2$/mean$^2$) in the 5--10 keV band, where the flux variabilities are
little affected by absorption. Thus, this frequency region is likely
below the low frequency break of the band-limited noise observed from
BHXBs in the low/hard state.

\subsection{Modeling time-averaged non-dip spectrum}

We extract the time-averaged XIS and HXD spectra in the deep dip,
shallow dip, and non-dip phases and analyze them separately. In this
subsection we concentrate on the non-dip spectra. We utilize the
energy bands of 0.7--9.0 keV, 0.7--8.0 keV, 12--70 keV, and 50--130 
keV for FI-XISs, BI-XIS, HXD/PIN, and HXD/GSO, respectively, where 
the signal-to-noise ratios are sufficiently good and the calibration 
is the most reliable.

\begin{figure*}
\plotone{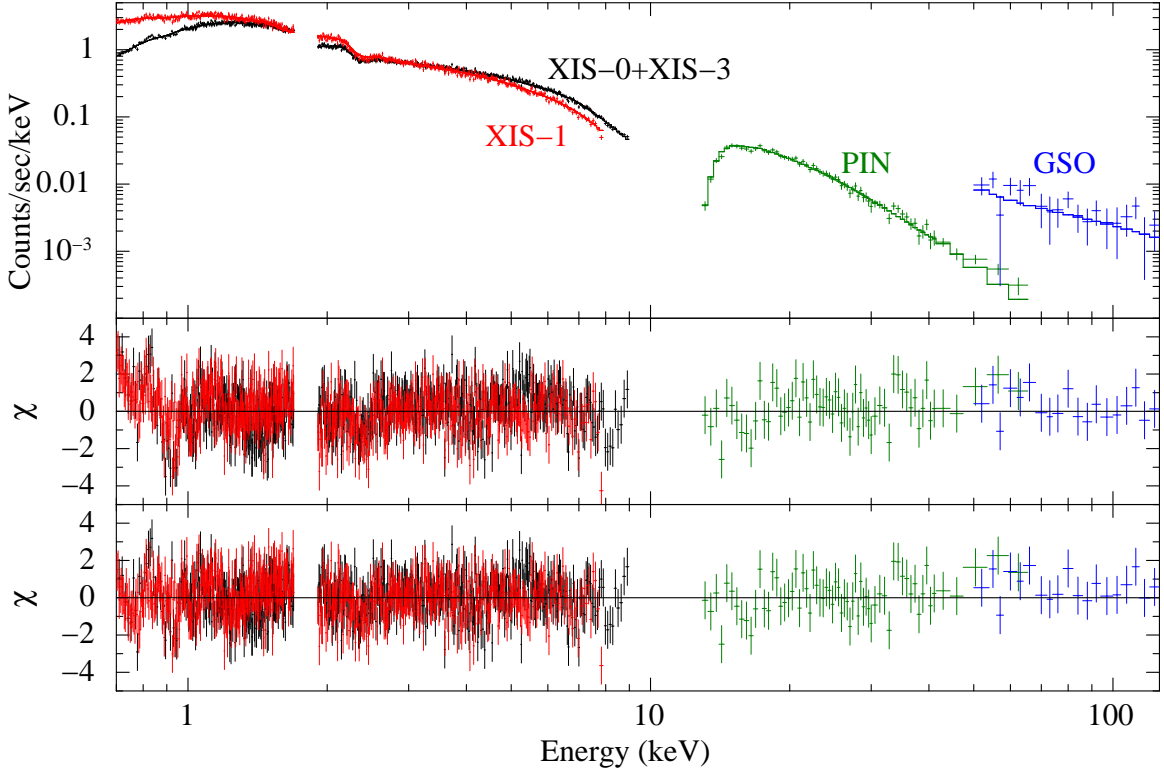}
\caption{The time-averaged {\it Suzaku} spectra in the non-dip phases fitted with 
a disk and a Comptonization components are plotted in the top panel. The middle 
and bottom panels show the residuals of the fits with the power-law model and 
the {\tt diskbb} $+$ {\tt nthcomp} model, respectively.
\label{fig_comp}}
\end{figure*}

\begin{figure}
\plotone{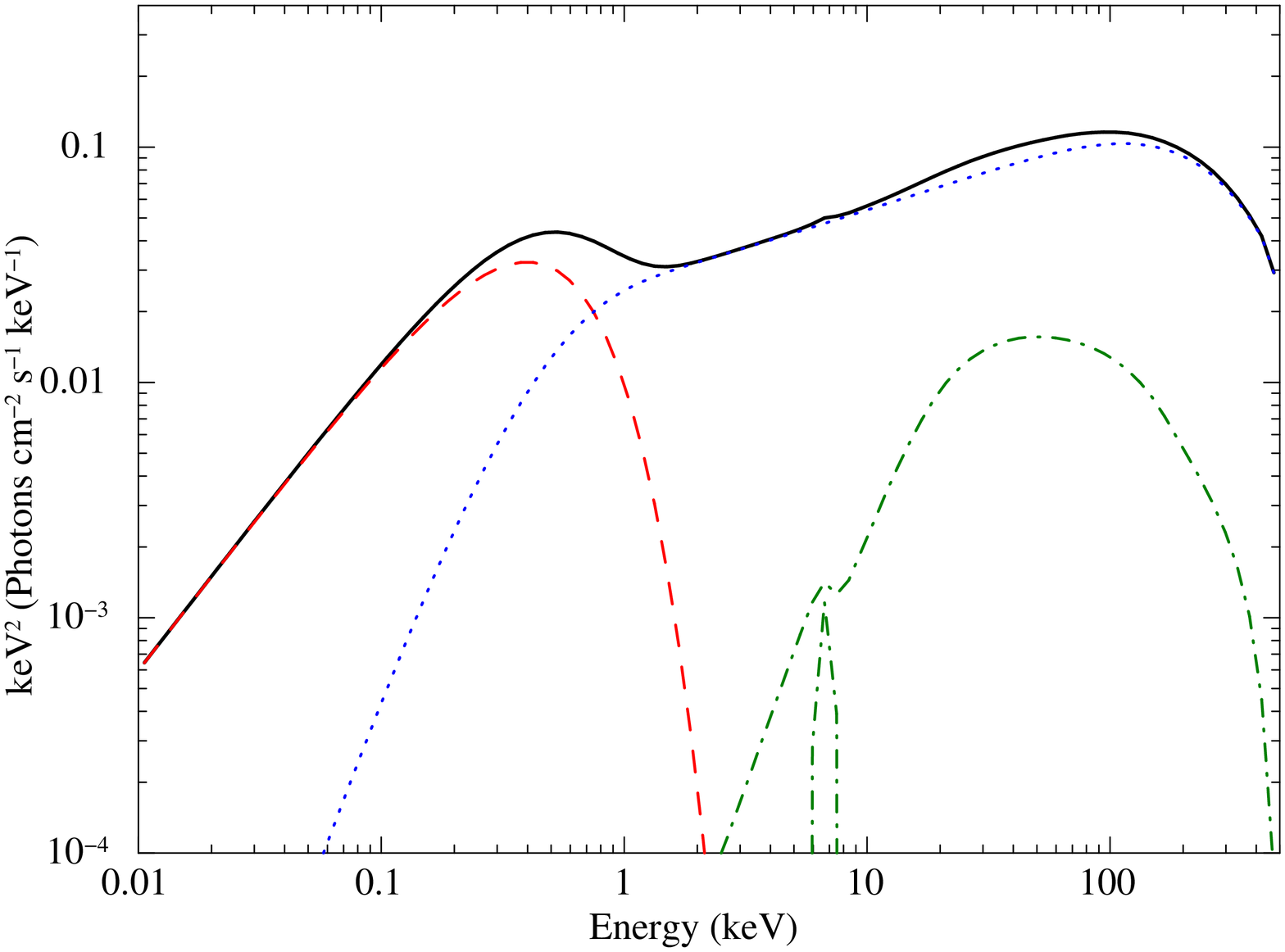}
\caption{The best-fit disk $+$ Comptonization spectrum (black, solid) and each component 
corrected for absorption are separately plotted in the $\nu F_{\nu}$ form. The red (dashed), 
blue (dotted), green (dash-dotted) lines represent the disk, Comptonization, reflection 
components, respectively.
\label{fig_comp2}}
\end{figure}

\begin{table*}
\begin{center}
\caption{The best-fit parameters of {\it Suzaku} spectra in the deep dip, shallow dip, and non-dip 
periods. \label{tab_comp}} \tabletypesize{\scriptsize} 
\begin{tabular}{llccc}
\tableline\tableline
Component & Parameter & non-dip & deep dip & shallow dip \\ \tableline
phabs & $N_{\rm H}$ ($10^{22}$ cm$^{-2}$) & $0.12 \pm 0.03$ & $0.23^{+0.05}_{-0.04}$ & $0.17 \pm 0.02$\\
xsabs & $N_{\rm H}$ ($10^{22}$ cm$^{-2}$) & $0.61^{+0.10}_{-0.09}$ & $14.4 \pm 0.6$ & $6.6^{+0.5}_{-0.4}$  \\
  & $\log \xi$ & $2.19 \pm 0.04 $ & $1.90 \pm 0.07$ & $1.79 \pm 0.07$ \\
  &blue shift (km)\tablenotemark{c} & $<2300$ & $<2700$ & $<5800$ \\
 & covering fraction & 1 (fix) & $0.91 \pm 0.01$& $0.72^{+0.03}_{-0.04}$ \\
diskbb & $kT_{\rm in} $ (keV) & $0.168^{+0.008}_{-0.006}$ & &  \\
 & norm & $6.0^{+3.4}_{-2.4} \times 10^{3}$  & &  \\
nthcomp & $\Gamma$ & $1.70^{+0.03}_{-0.02}$ & & \\
 & $E_{\rm cut}$ (keV)& $300$ (fix) & & \\
 & norm & $2.46^{+0.09}_{-0.08} \times 10^{-2}$ & & \\
reflect & $\Omega/2\pi$ & $0.4 \pm 0.2$ & & \\
 & $i$ (deg)& 75 (fix) & & \\
gauss \tablenotemark{a} & $E_{\rm cen}$ (keV) & 
6.4 (fix) & & \\
& $\sigma$ (eV) & 10 (fix) & &  \\
\tableline
 $\chi^2/{\rm d.o.f.}$ & & $1269/1165$ & $540/535$ & $1070/994$ \\
flux \footnotemark{b} & & $1.3\times10^{-10}$  
& $7.1 \times10^{-11}$ & $1.0\times10^{-10}$ \\
\tableline
\end{tabular}
\tablenotetext{a}{The normalization of Gaussian component is linked to the reflection 
strength $\Omega/2\pi$ of the {\tt reflect} model so that the equivalent width with 
respect to the reflection continuum is $\approx$ 1.0 keV.}
\tablenotetext{b}{absorbed 1--10 keV flux (ergs cm$^{-2}$ sec$^{-1}$)}
\tablenotetext{c}{Positive values represent blue shifts.}
\tablecomments{The non-dip spectrum is fitted with 
{\tt phabs*xsabs*(diskbb+kdblur*(reflect*nthcomp+gauss))}, 
where {\tt xsabs} is an ionized absorption model created with XSTAR. 
In fitting the two dip spectra, all the parameters except for those 
of the neutral and ionized absorption components are fixed at the best-fit 
values of the non-dip spectrum. The blank columns in the table 
of the dip spectra are the fixed parameters. Partial covering of the ionized 
absorber is included in the model of dipping spectra.}
\end{center}
\end{table*}

The non-dip spectrum is roughly characterized by a power-law component
extended up to 130 keV with a photon index of $\approx 1.591 \pm
0.005$ ($\chi^2/{\rm d.o.f.} = 1472/1171$), although we find broad 
depressions around 0.75 keV and 0.9 keV. 
These structures likely correspond to the photoelectronic
absorption lines and/or edges of highly ionized oxygen-K and iron-L
shells, which are similar to ``warm absorbers'' seen in many active
galactic nuclei like MCG-6--30--15 \citep[e.g.,][]{nan92,fab94} and
NGC~4051 \citep[e.g.,][]{pou94, mih94}. The hard spectral shape
suggests that the source stayed in the low/hard state during our
observation. The exponential cutoff is not detected within the energy
band of the non-dip spectrum ($<130$ keV), and therefore in the 
following analysis, we fix the cutoff energy at 300 keV, which is 
within the typical value observed from BHXBs in the low/hard state 
\citep[$\lesssim 300$ keV; see e.g.,][]{tan96}.

To investigate the detailed properties of the accretion flow and the
ionized absorber, we next analyze the non-dip spectrum with a more
sophisticated model. Following the general description of the X-ray
spectrum in the low/hard state \citep[e.g.,][]{gie97}, we adopt a model 
composed of the multicolor disk (MCD) emission and its thermal 
Comptonization. The {\tt nthcomp} model \citep{zdz96,zyc99} and the 
{\tt diskbb} model \citep{mit84} are employed to represent the 
Comptonization and direct emission from the disk, respectively. We assume 
that all the seed photons for the Comptonized component are produced by 
the disk, and link the seed temperature of the {\tt nthcomp} model 
to the inner disk temperature of the MCD component. 
We add {\tt phabs} as interstellar absorption, assuming the solar
abundance. To consider reflection of Comptonized photons on the disk,
we convolve the {\tt nthcomp} component with the {\tt reflect}
model. This model calculates a reflected spectrum from neutral material
\citep{mag95}\footnote{We examine the ionization level of the
reflector utilizing the {\tt ireflect} model, an ionized version of
{\tt reflect}, but the ionization parameter ($\xi$) is not constrained
at all. We obtain the minimum reduced chi-squared in the case of a
neutral reflector ($\xi = 0$).}. The {\tt reflect} model does not
contain the iron K$\alpha$ emission line, whose equivalent width is
$\approx 1$ keV with respect to the reflected continuum, as suggested
by numerical calculations \citep[e.g.,][]{mat91}. Hence, we add a
Gaussian component as the iron-K$\alpha$ emission line and fix the
line energy and the line width at 6.4 keV and 10 eV, respectively.
The normalization of Gaussian component is linked to the reflection 
strength $\Omega/2\pi$ of the {\tt reflect} model so that the equivalent 
width with respect to the reflection continuum is $\approx$ 1 keV 
\citep[e.g.,][]{mat91}\footnote{We first fitted with the strength of the Gaussian 
unlinked with that of the reflection continuum. However, the upper limit of 
the line flux was found to be unreasonably small, an order of magnitude 
lower than what is expected from the equivalent width with respect to a 
reflection continuum ($\approx$ 1 keV). }.
The reflection spectrum is smeared with {\tt kdblur}, which calculates
relativistic effects from an accretion disk around a rotating black
hole using the results of \citet{lao91}. We assume a inclination angle
of 75$^\circ$ (see Section 5.1) and the index for the radial
dependence of emissivity ($\beta$, where emissivity $\propto
r^{-\beta}$) of $3$, and an outer radius of 400 $R_{\rm g}$ ($R_{\rm
g}$ represents the gravitational radius, $GM_{\rm BH}/c^2$). 
The inner radius is first left as a free parameter but it is totally 
unconstrained. We therefore fix the inner radius at 100 $R_{\rm g}$ in 
the following spectral fits.  The energy range for the model calculation 
is extended to 0.1--1000 keV to apply the convolution models.

To analyze the ionized absorption features, we create a photoionized
absorption model with the spectral synthesis code XSTAR version
2.2.1bk, assuming that the ionized absorber has the solar abundances and
that its turbulent velocity is 300 km sec$^{-1}$. This model can be 
used in XSPEC as an multiplicative component with free parameters
of the equivalent hydrogen column density ($N_{\rm H}$), ionization
parameter ($\xi = L_{\rm X}/n_{\rm H}R^2$, where $L_{\rm X}$, $n_{\rm
H}$, and $R$ represent the ionizing flux in the energy range of
1--1000 Ry (Rydberg unit; 1 Ry $=$ 13.6 eV), the number density 
of hydrogen nuclei, and the distance from the 
X-ray source to absorber, respectively), and the Doppler
shift. The absorption spectra calculated with XSTAR depend on the
spectral shape of the incident radiation on the absorber. We first
adopt a single power-law model with a photon index of 1.6 as the input
spectrum for XSTAR and fit the resulting model to the non-dip
spectrum. Next, we re-create an XSTAR absorption model using the
best-fit unabsorbed continuum model and then refit the spectrum with
the newly obtained absorption model. These steps are performed in an
iterative manner, until the parameters of the continuum model 
become identical with those obtained in the previous iteration 
within the ranges of 90\% errors. 
In the following, we show the final best-fit results after the iteration.

The final fitting model is thus expressed as {\tt
phabs*\\xsabs*(diskbb+kdblur*(gaussian+reflect*nthcomp))}, where {\tt
xsabs} represents the XSTAR ionized absorption model. The spectra and
the best-fit model are shown in Figure~\ref{fig_comp} and 
Figure~\ref{fig_comp2}, and the
resulting parameters are given in Table~\ref{tab_comp}. 
We find that this model describes the {\it Suzaku} spectra reasonably well, 
with $\chi^2/{\rm d.o.f.} = 1269/1165$. The fit quality is improved 
from that of the power-law model with an F-test probability of 
$1 \times 10^{-34}$. The ionization parameter and column
density are estimated as $\log \xi = 2.19 \pm 0.04$ and $N_{\rm
H} = (6.1^{+1.0}_{-0.9}) \times 10^{21}$ cm$^{-2}$ for the ionized
absorber, respectively. The Doppler shift is not 
detected with an upper limit of $< 2300$ km sec$^{-1}$.  A small inner
disk temperature ($0.168^{+0.008}_{-0.006}$ keV) and a large
normalization of the MCD model ($6.0^{+3.4}_{-2.4} \times 10^{3}$) are
obtained, which suggest that the standard disk is truncated (see
Section 5.3) during the {\it Suzaku} observation. The hydrogen column
density of neutral absorption, $(1.2 \pm 0.3) \times 10^{21}$ cm$^{-2}$,
is comparable to the total Galactic column in the
direction of MAXI~J1305$-$704 ($\approx 1.8 \times 10^{21}$ cm$^{-2}$),
estimated from the HI all-sky map by \citet{kal05} by utilizing the 
{\tt nh} ftool.

\begin{figure}
\plotone{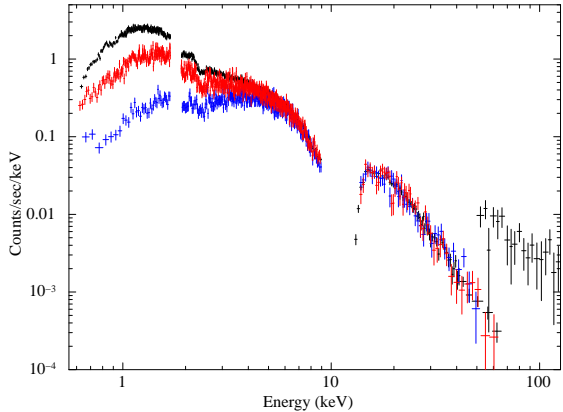}
\caption{The time-averaged {\it Suzaku} spectra in the deep dip (blue, hardest one) and 
the shallow dip (red, second hardest one), compared with that in the non-dip 
period (black). The XIS-1 spectra are not shown for clarity.\label{fig_specall}}
\end{figure}

Recent {\it Suzaku} observation of BHXBs in the low/hard state have revealed
that the Comptonized plasmas are more complex than a single-zone
homogeneous structure. \citet{tak08}, \citet{mak08}, and \citet{shi11}
reproduced the time-averaged spectra with double Comptonization
components that have different optical depths.  Furthermore,
\citet{yam13} successfully separated the second variable component
from the {\it Suzaku} spectra of Cyg X-1 in the low/hard state by
considering timing information. Here we investigate whether or not
these complex structures are also detected in MAXI~J1305$-$704. We add
another {\tt nthcomp} component to the single {\tt nthcomp} model to
consider the double Comptonization corona. The seed temperatures of
the two {\tt nthcomp} models are linked to the inner disk temperature
of the MCD component. We find, however, that this 
``double {\tt nthcomp}'' model does not improve the fit.

\subsection{Analysis of Dip spectra}

\begin{figure}
\plotone{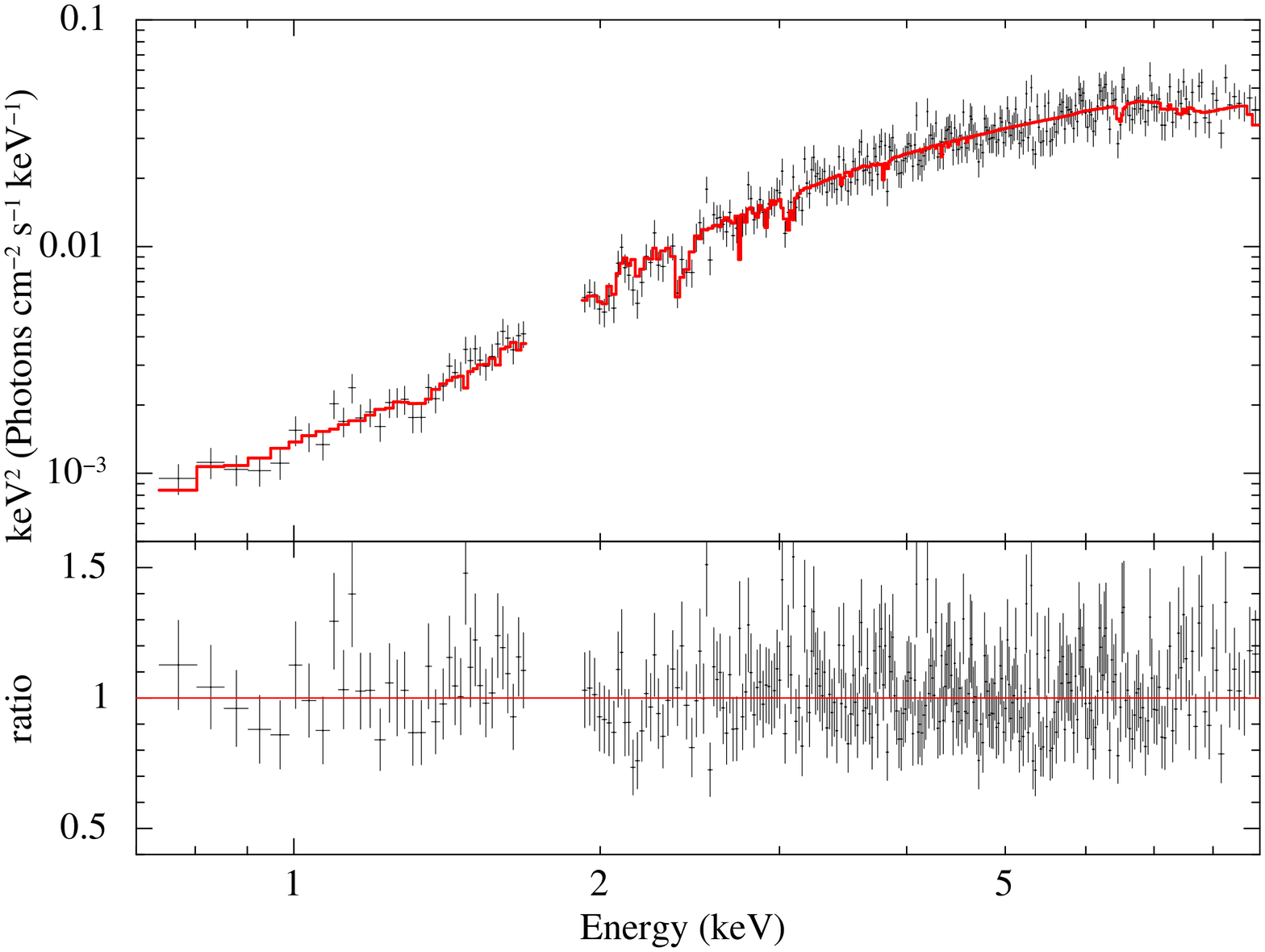}
\plotone{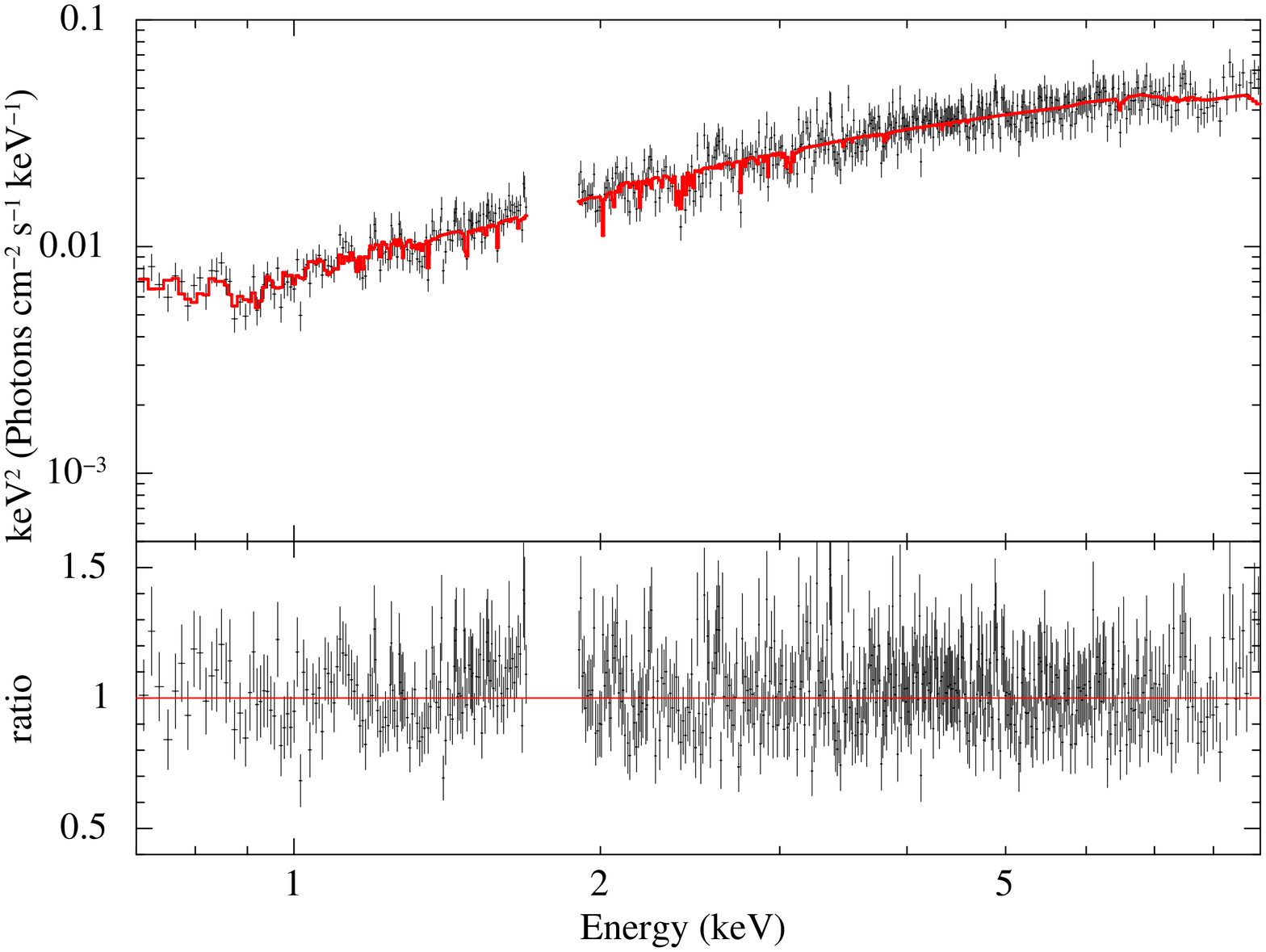}
\plotone{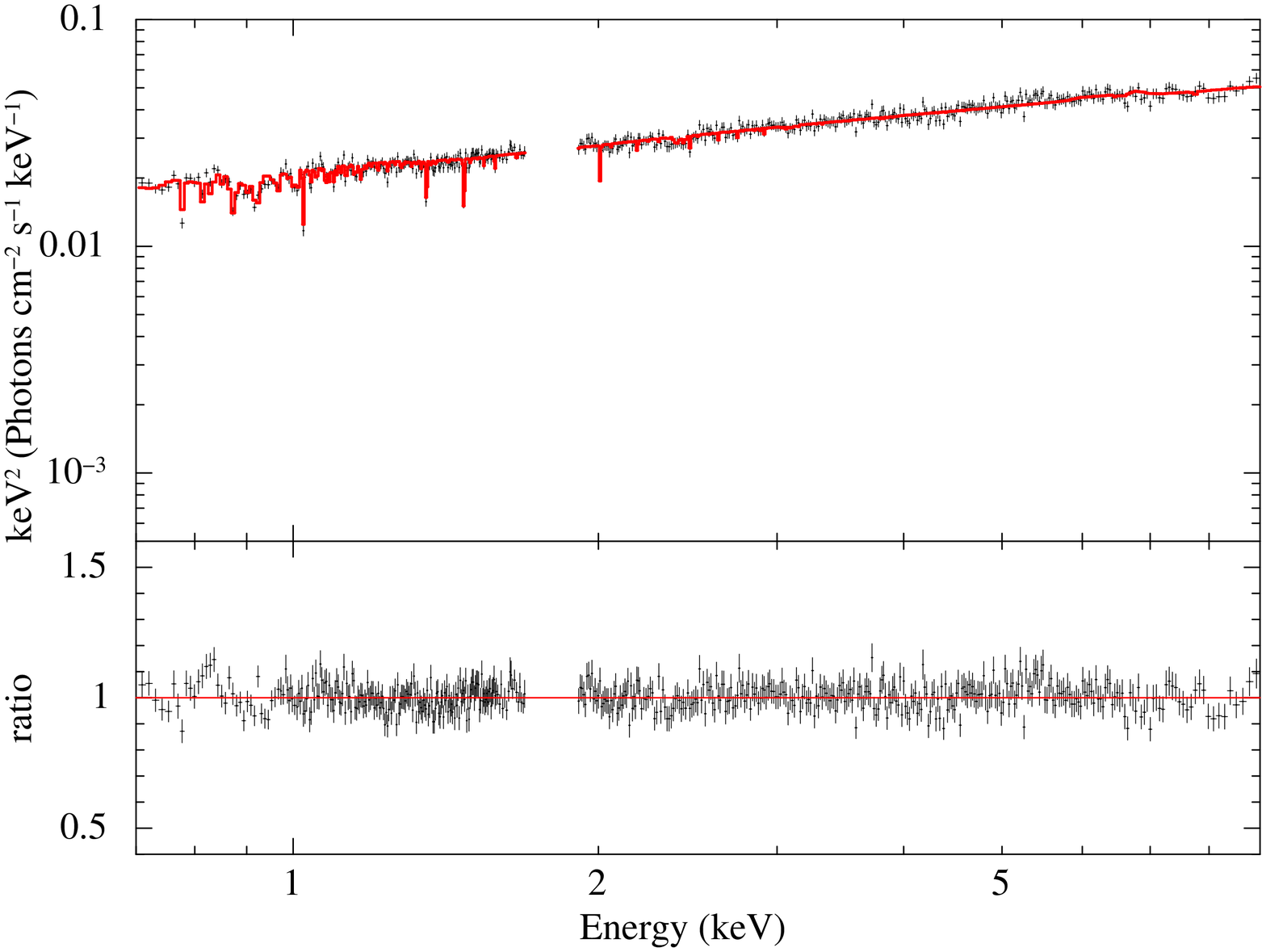}
\caption{The spectra, best-fit models, and data vs. model ratios 
in the dipping and non-dip phases. The top, middle, and bottom 
panels show the results in the deep dip, shallow dip, and 
non-dip phases, respectively. The dipping spectra are fitted with 
the best-fit model of the non-dip spectrum. 
The XIS-1 and HXD spectra are ignored in all panels for 
illustrating purposes.
\label{fig_specdip}}
\end{figure}

\begin{figure*}
\plottwo{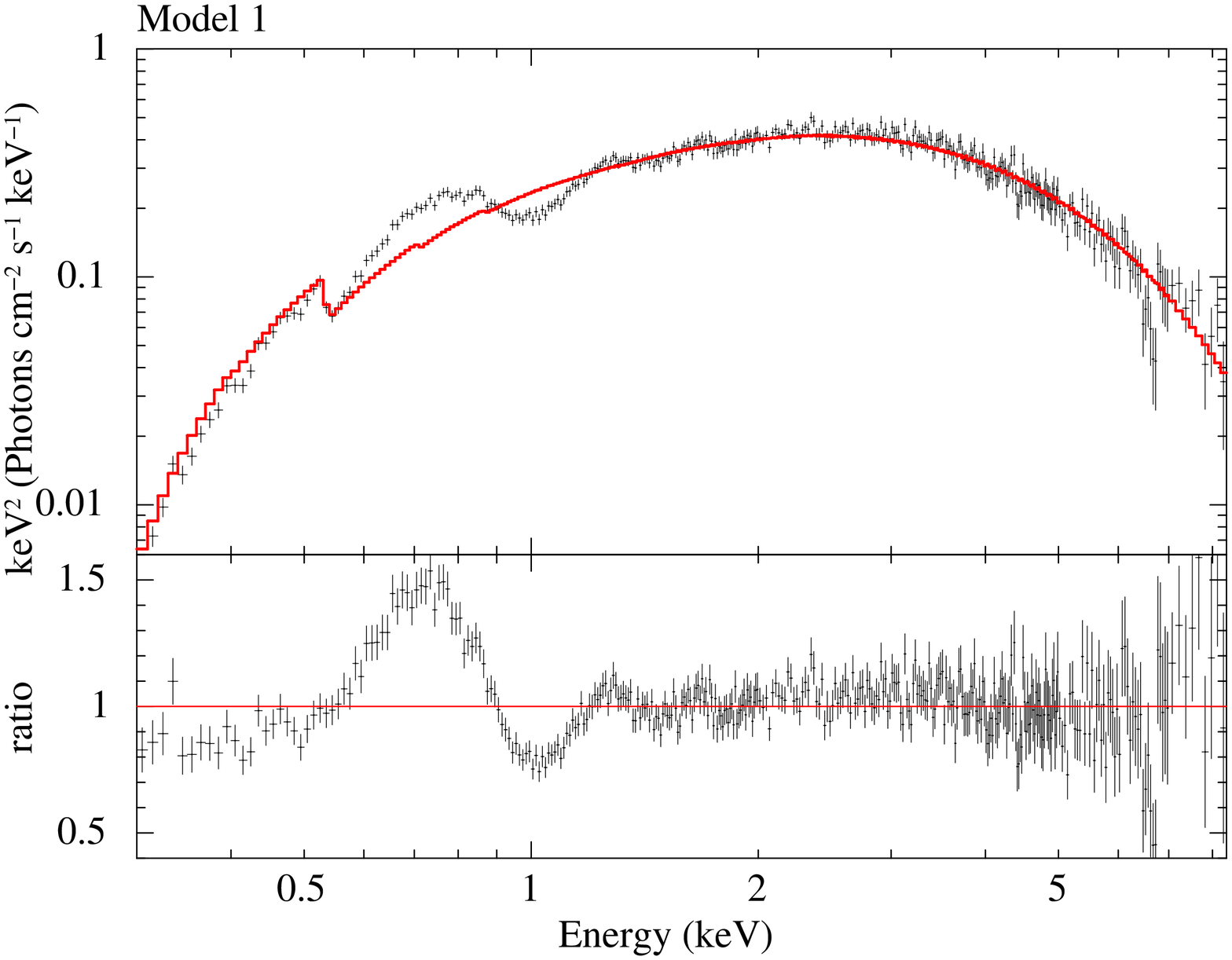}{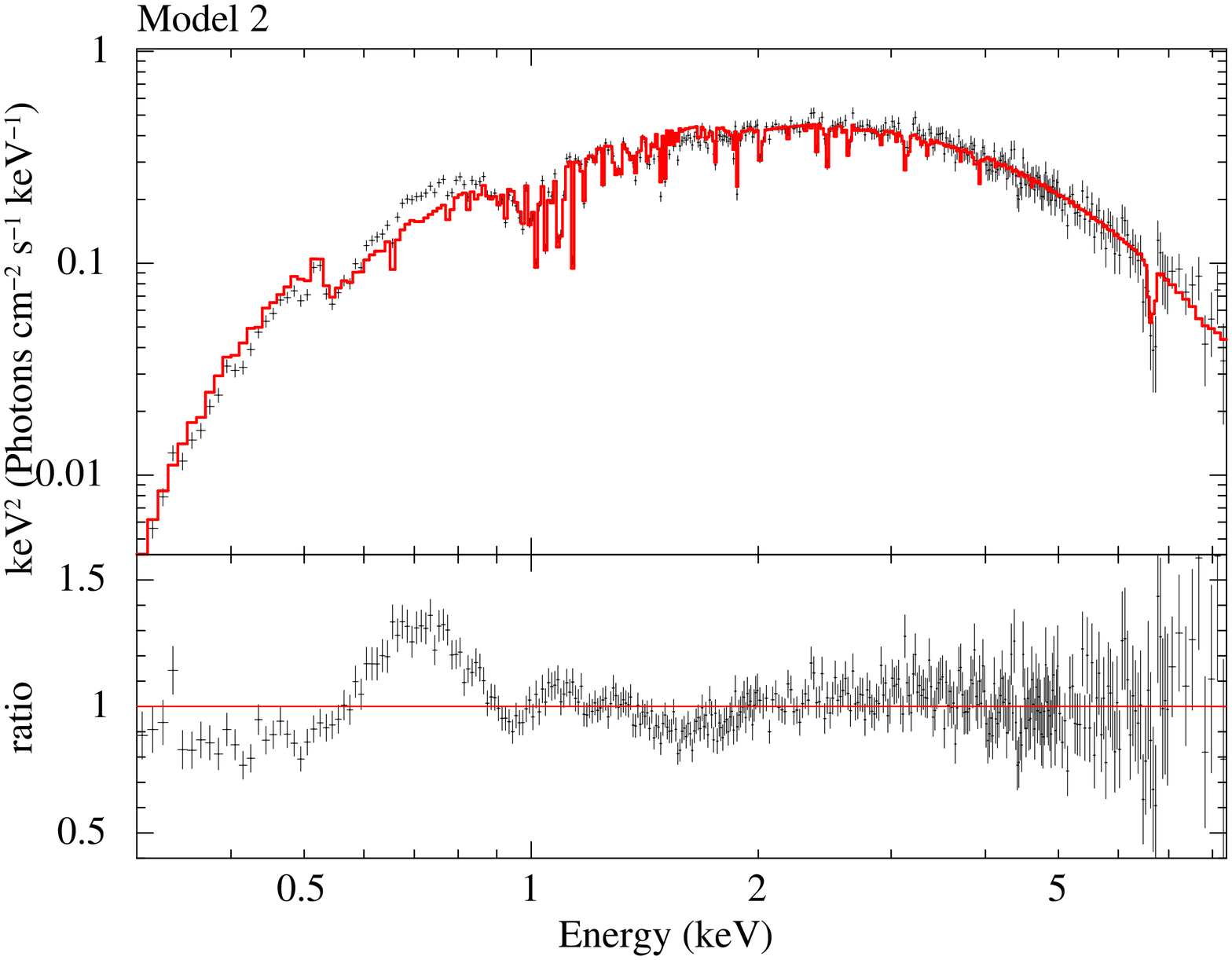}
\plottwo{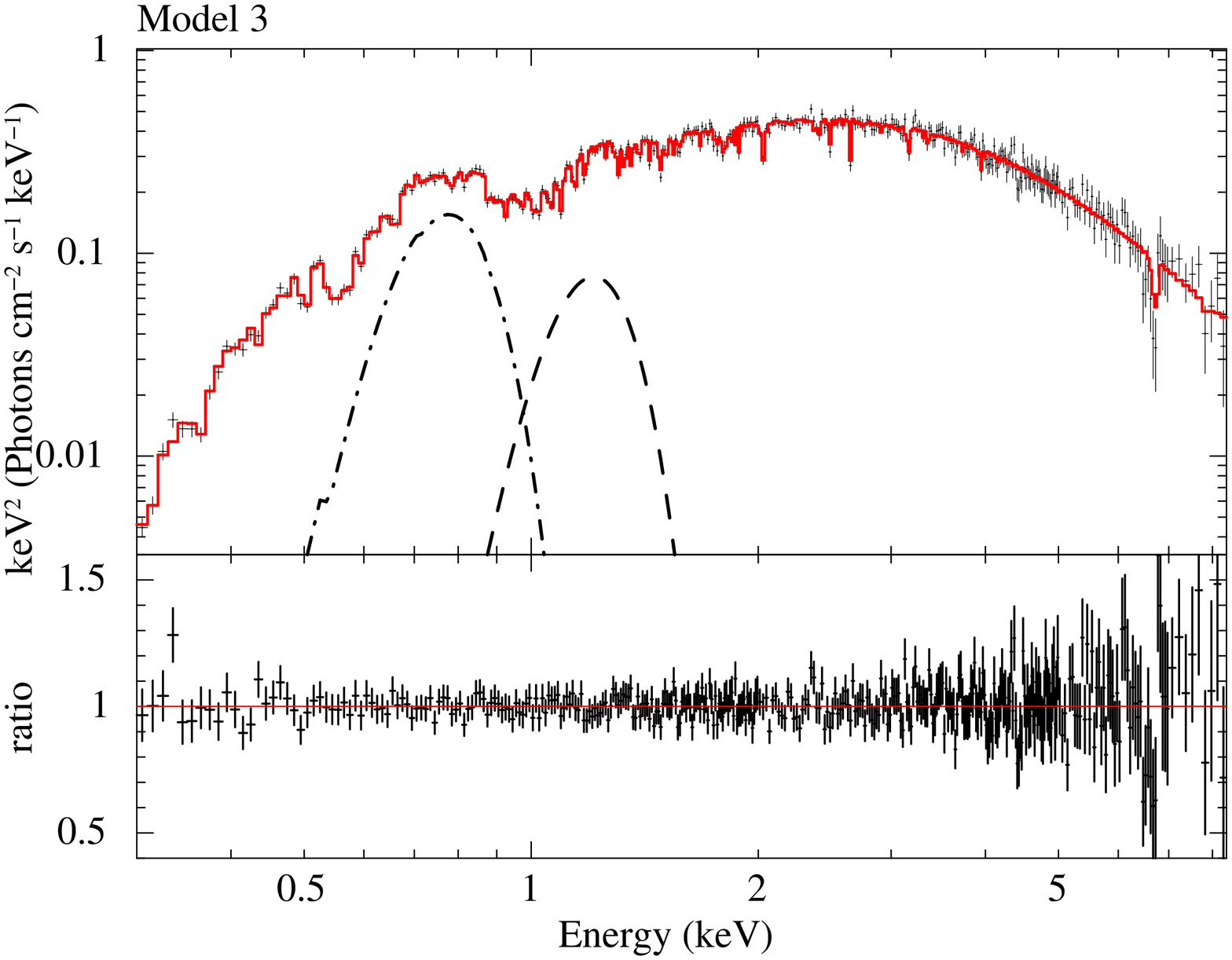}{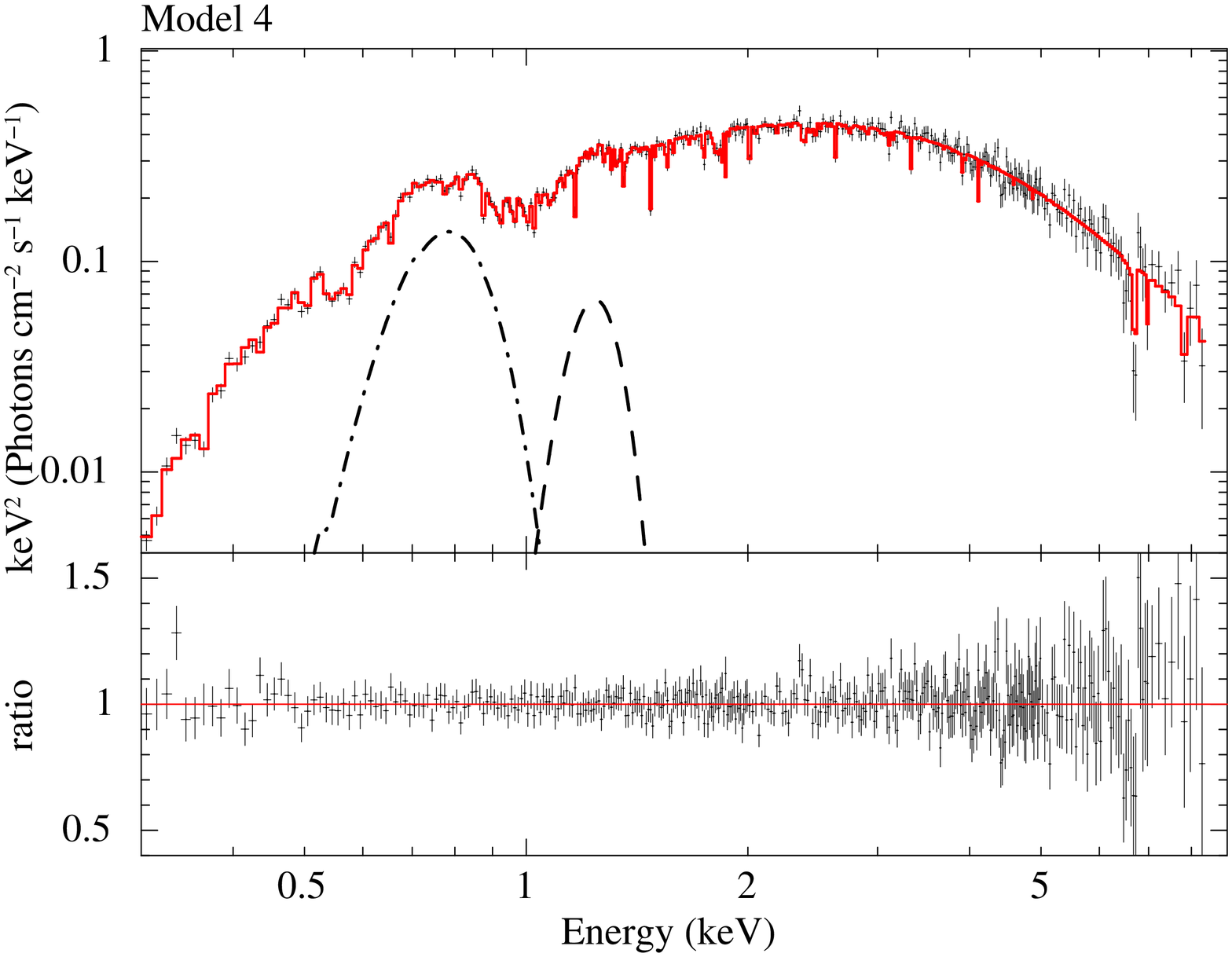}
\caption{The time-averaged spectrum of XRT spectrum fitted with various models. 
The lower panel represents the data vs. model ratios in each bin. 
(Top left) {\tt phabs*simpl*bhspec}. (Top right) {\tt phabs*xsabs1*simpl*bhspec}. 
(Bottom left) {\tt phabs*(xsabs1*xsabs2*simpl*bhspec$+$gauss$+$gauss)}. 
(Bottom right) {\tt phabs*(xsabs1*xsabs2*simpl*diskbb$+$gauss$+$gauss)}, where 
``xsabs'' is the ionized absorption model created with XSTAR.\label{fig_xrt}}
\end{figure*}

We analyze the deep and shallow dip spectra with the same model that
used for the non-dip spectrum. Figure~\ref{fig_specall} compares the
XIS and HXD spectra in the deep dip, shallow dip, and non-dip
phases. We obtain the HXD/PIN spectrum up to 60 keV for the shallow
dip and to 50 keV for the deep dip. The HXD/GSO data in the dips are
not usable due to the limited photon statistics, however. We employ
the final results described in the previous section and fix all the parameter 
at the best-fit value of the non-dip spectrum, except for those of the 
neutral and ionized absorption components. Considering
that dipping spectra often modelled with partial absorbers, we
introduce a covering fraction of the ionized absorber. The total
fitting model for the dip spectra is described as {\tt
phabs*(f*xsabs+(1-f))*(diskbb+kdblur*(gaussian\\+reflect*nthcomp))},
where {\tt f} corresponds to the covering fraction.

This model successfully reproduce the dip spectra, yielding
$\chi^2/{\rm d.o.f.} = 540/535$, and $1070/994$ for the deep and
shallow dips, respectively. We find that the two dip spectra can
be described with more than one order of magnitude larger column
densities and about a factor of 2 smaller ionization parameters than
those of the non-dip spectrum. The column density of the deep dip is twice
as much as that of shallow dip.  The covering fraction $f$ is
estimated as $0.72^{+0.03}_{-0.04}$ for the shallow dip, while the deep dip
spectrum is almost totally absorbed, with $f=0.91 \pm 0.01$. The
resulting parameters are listed in Table~\ref{tab_comp} and the
best-fit spectra are plotted in Figure~\ref{fig_specdip}.

The xsabs model do not include the Comptonization in the ionized 
absorber itself. This might affect the fits particularly for the deep dip 
spectrum, in which the ionized absorber has relatively large column 
density ($N_{\rm H} \sim 10^{23}$ cm$^{-1}$). To account for the possible 
effects of Compton scattering, we add the {\tt cabs} model to the final 
model with its column density linked to that of the ionized absorber, 
and re-fit the deep dip spectrum. We find, however, that the 
effects are negligible and all the parameters remain unchanged within 
their 90\% confidence ranges. 
We confirm that the energy dependence of the scattering cross section,
which is not included in {\tt cabs}, is also negligible in our energy
range. The column density of the deep dip corresponds to an optical
depth of $\tau \approx 0.1$ in Thomson scattering, which only reduces
to $\tau \approx 0.08$ at 50 keV.

\subsection{{\it Swift}/XRT spectrum in the high/soft state}

\begin{table*}
\begin{center}
\caption{The fitting results of {\it Swift}/XRT non-dip spectrum.\label{tab_xrt}}
\begin{tabular}{llccccc}
\tableline\tableline
Component & Parameter & Model1\tablenotemark{a} 
& Model2\tablenotemark{b}
& Model3\tablenotemark{c}
& Model4\tablenotemark{d}  \\
\tableline
phabs & $N_{\rm H}$ ($10^{22}$ cm$^{-2}$) & $0.097 \pm 0.002$  &  $0.123 \pm 0.004$ 
 & $0.097^{+0.020}_{-0.018}$ & $0.098 \pm 0.006$\\
xsabs1\tablenotemark{e} 
 & $N_{\rm H}$ ($10^{22}$ cm$^{-2}$) & --  
 & $7.8^{+1.4}_{-1.3}$  & $5.1^{+3.8}_{-2.8}$ & $5.8^{+41.7}_{-3.4}$ \\
  & $\log \xi$ & -- & $2.64 \pm 0.06$ 
& $2.86^{+0.52}_{-0.18}$ & $3.13_{-0.26}^{+1.87, {\rm pegged}}$ \\
  &blue shift (km)\tablenotemark{f} & -- & $1700^{+1000}_{-1200}$ & 
  $<4800$ & $<4300$\\ 
 xsabs2\tablenotemark{e} 
 & $N_{\rm H}$ ($10^{22}$ cm$^{-2}$)  & --  
 & -- & $1.0^{+0.4}_{-0.3}$ & $1.5^{+0.6}_{-0.5}$  \\
  & $\log \xi$ & -- & -- 
  & $1.2 \pm 0.2$ & $1.7 \pm 0.1$\\
  &blue shift (km)\tablenotemark{f} & -- & -- & 0 (fix) & 0 (fix) \\
diskbb & $T_{\rm in}$ (keV)  & -- & -- & -- & $0.88^{+0.03}_{-0.04}$ \\
 & norm  & -- & --  & -- & $139^{+26}_{-45}$\\
bhspec\tablenotemark{g}  
& $a$ & $0.69 \pm 0.01$ & $0.56 \pm 0.02$ & $0.46 \pm 0.06$& --\\
 & norm & $1.77^{+0.01}_{-0.02}$ & $2.35 \pm 0.06$ & $2.5^{+0.3}_{-0.2}$&--  \\
simpl & $\Gamma$  & $2.2$ (fix) & $2.2$ (fix) & $2.2$ (fix) & $2.2$ (fix)   \\
  & scattering fraction & $<0.003$ & $0.02 \pm 0.01$  & $0.03 \pm 0.01$ & $0.04 \pm 0.01$\\
gauss 
& line energy (keV) & -- & -- & $1.17^{+0.03}_{-0.04} $&  $1.18^{+0.06}_{-0.05} $\\
& $\sigma$ & & & $0.14 \pm 0.03$ & $0.12^{+0.04}_{-0.07} $ \\
  & norm &  -- & -- & $0.022^{+0.009}_{-0.008} $ & $0.017^{+0.006}_{-0.008} $ \\
& E.W. (eV) & -- & -- & $61^{+35}_{-22}$ &$53^{+32}_{-29}$  \\
gauss
& line energy (keV) & -- & -- & $0.74 \pm 0.01 $&  $0.75 \pm 0.01 $\\
& $\sigma$ & & & $0.10 \pm 0.01$ & $0.10 \pm 0.01$ \\
  & norm & -- & -- & $0.11 \pm 0.01$ & $0.10 \pm 0.01$ \\
& E.W. (eV) & -- & -- & $159^{+22}_{-14}$& $183^{+18}_{-17}$  \\
\tableline
$\chi^2/{\rm d.o.f}$ & & $1687/370$  & $1018/367$ & $294/359$ & $292/359$ \\
\tableline
\end{tabular}
\tablenotetext{a}{\tt phabs*simpl*bhspec}
\tablenotetext{b}{\tt phabs*xsabs1*simpl*bhspec} 
\tablenotetext{c}{\tt phabs*(xsabs2*xsabs1*simpl*bhspec$+$gauss$+$gauss)} 
\tablenotetext{d}{{\tt phabs*(xsabs2*xsabs1*simpl*diskbb$+$gauss$+$gauss)}, for 
direct comparison with the disk flux obtained from the {\it Suzaku} non-dip spectrum.}
\tablenotetext{e}{Ionized absorption model created with XSTAR. Incident spectrum 
is defined as the {\tt diskbb} model with a inner temperature of 1.0 keV. We assume a 
turbulent velocity of 300 km sec$^{-1}$.}
\tablenotetext{f}{Positive values represent blue shifts.}
\tablenotetext{g}{We assume $i = 75^\circ$, $M_{\rm BH} = 3 M_{\sun}$, 
luminosity $= 0.05 L_{\rm Edd}$, and $\alpha = 0.01$, where $\alpha$ represents the 
viscosity parameter in the \citet{sha73} prescription for the stress 
$\tau_{r\phi}= \alpha \times P$ ($P$ is the total pressure). }
\end{center}
\end{table*}

To compare the {\it Suzaku} non-dip spectrum in the low/hard state with 
spectra in the high/soft state, we
analyze a {\it Swift}/XRT spectrum of MAXI~J1305$-$704 obtained from 2012
April 19 to 21 during the high/soft state. As described in
\citet{mil12a}, these XRT data also show dipping behaviors. We create
a time-averaged non-dip spectrum by extracting the events when the count
rate exceeds 20 counts sec$^{-1}$ in the 1--10 keV light curve with 16
sec bins.  Following the release note for the {\it Swift} XRT
CALDB\footnote{http://heasarc.gsfc.nasa.gov/docs/heasarc/caldb/swift/docs/\\xrt/SWIFT-XRT-CALDB-09\_v16.pdf},
we use the data down to 0.3 keV for the following spectral fit, where
the calibration of the energy response is reliable.

First we fit the XRT spectrum using an MCD model with a neutral absorption. 
A {\tt simpl} model \citep{ste09} is also incorporated to account for 
Comptonization, with a fixed photon index of 2.2, a typical value in 
the high/soft state of BHXBs \citep[e.g.,][]{ebi94,don07,kol11}. We extend the energy 
range to 0.01--100 keV for the model calculation since {\tt simpl} is a 
convolution model. We find that this model, {\tt phabs*simpl*diskbb}, roughly
describes the XRT spectrum, with an inner disk temperature ($\approx 1.0$ keV) 
and a small scattering fraction ($< 2.6$\% of the total disk emission), 
although the fit is far from acceptable ($\chi^2/{\rm d.o.f.}
= 1749/370$) mainly due to the broad absorption (and/or emission)-like
structures in the soft band below $\approx$ 1 keV. These large residuals
are probably a composition of the iron-L absorption lines, which are
detected in the {\it Chandra} HETGS observation on 10 days after the 
{\it Swift} observation, as reported by \citet{mil12b}. The XRT spectrum also 
has a narrow absorption line at about 6.6 keV, which likely corresponds to
K-$\alpha$ lines of highly ionized iron ions. By fitting the line with
a negative Gaussian, its center energy, line flux, and equivalent width 
are estimated to be $6.57^{+0.09}_{-0.08}$ keV, $2.7^{+1.4}_{-1.1} \times 10^{-4}$ 
ergs cm$^{-2}$ sec$^{-2}$, and $49^{+27}_{-23}$ eV, respectively.

We find that the unabsorbed flux in the 0.01--100 keV band, $1.4 \times
10^{-9}$ erg cm$^{-2}$ sec $^{-1}$, is only $\approx 2.3$ times larger
than that of the {\it Suzaku} non-dip spectrum ($6.1 \times 10^{-10}$ erg
cm$^{-2}$ sec $^{-1}$). This suggests that MAXI~J1305$-$704 was in a
relatively faint high/soft state and that the bolometric luminosity in the 
{\it Swift} observation was comparable with that in the soft-to-hard 
transition, typically $\approx 0.02 L_{\rm Edd}$ \citep{mcc03}. 
However, the inner disk temperature is too high to be expected for such 
a faint high/soft state with a low accretion
rate. This could be understood in the way that the strong relativistic
beaming effects due to a high inclination angle and/or a high black hole
spin significantly modify the disk spectrum and consequently we obtain
an apparently higher inner temperature than the intrinsic one. We therefore
replace {\tt diskbb} with {\tt bhspec} \citep{dav05}, a relativistic
disk emission model, to fit the spectra (i.e., {\tt phabs*simpl*bhspec}; 
Model 1 in Table~\ref{tab_xrt}). The {\tt bhspec} model
calculates the radiation transfer in the accretion disk around a black
hole by self-consistently considering its vertical structure. 
The model parameters are the black hole mass ($M_{\rm BH}$), spin
parameter ($a = cJ/GM_{\rm BH}^2$, where $J$ represents angular momentum
of the black hole), distance, inclination angle, disk luminosity, and
the $\alpha$ parameter, which we fix at 0.01. Here we assume a black hole
mass of $M_{\rm BH} = 3 M_\sun$, a high inclination, $i = 75^\circ$ (see
Section 5.1), and a disk luminosity corresponding to $0.05 L_{\rm Edd}$,
and leave $a$ and the normalization $K$, which is related to the 
distance $d$ via $K =$ (10 kpc/$d$)$^{2}$, as free parameters. We find 
that the {\it Swift}/XRT spectrum favors this {\tt bhspec} model better 
than the {\tt diskbb} model with a smaller reduced chi-squared value
($\chi^2/{\rm d.o.f.} = 1687/370$). A moderate spin parameter ($a \approx
0.7$) is obtained.

To investigate the properties of the ionized absorber responsible for
the structures at 6.6 keV and below $\approx$ 1 keV, we create a
multiplicative photoionized absorption model by utilizing XSTAR to fit
the non-dip XRT spectrum. We adopt an MCD with an inner temperature of
1.0 keV as the incident spectrum in the energy range of 1--1000 Ry. The
ionized absorber is assumed to have the solar abundances and a turbulent
velocity of 300 km sec$^{-1}$. The fit is much improved ($\chi^2/{\rm
d.o.f.} = 1018/367$) by using this model, {\tt phabs*xsabs*simpl*bhspec} 
(Model 2 in Table~\ref{tab_xrt}), where the {\tt xsabs} represents the
photoionized absorption. 
We obtain an ionization parameter of $\log \xi \approx 2.6$ and an 
equivalent hydrogen column density of $\approx 8 \times 10^{22}$ 
cm$^{-2}$.

The fit is still far from acceptable, however, due to the large 
residuals at around 0.7 keV and 1.2 keV, which cannot be explained by
calibration uncertainties of the response. The iron-K absorption line 
is not well modeled either, likely because these large structures at 
around 1 keV lead to wrong constraints on the parameters of the 
ionized absorption component.
The residuals are not originated from the uncertainties of {\tt bhspec} 
in modelling absorption edges, either. We confirm that the quality of fit is 
not improved by replacing the {\tt bhspec} component with another
relativistic disk emission model, {\tt kerrbb}, which does not consider
the vertical structure of the disk and has no absorption edges. 
Furthermore, they are reproduced neither by ionized O and Ne 
edges of additional absorption components nor by a superposition of 
emission lines created by the ionized absorber, 
which can never produce huge equivalent widths to fit the structures. 
We also change the oxygen, neon and iron abundances in the model, 
which could produce artificial emission and/or absorption-like structures 
around 0.5--1.0 keV if not appropriate. While the fit is not improved by 
varying the iron abundance, a better fit is obtained with a smaller oxygen 
abundance: $\chi^2/{\rm d.o.f.} = 734/367$ for an oxygen abundance of 0.5 
in the solar unit both for neutral and ionized absorbers. Nevertheless, we 
find that huge residuals still remain at $\approx$ 0.7 keV even in
the extreme case of no oxygen. A larger neon abundance also gives  
a better fit but the improvement of chi-squared value is not so significant 
as the oxygen abundance (at best $\chi^2/{\rm d.o.f.} = 930/366$ for a 
neon abundance of 9.0 in the solar unit). Thus, the oxygen and neon 
abundances in the absorbers cannot entirely explain the differences 
between the data and model in the soft energy band.

We find that the structures below $\approx$ 1 keV are well reproduced by
empirically adding two broad Gaussian components at 0.75 keV and 1.2 
keV with line widths of $\approx 100$ eV and 140 eV, and equivalent
widths of $\approx 160$ eV and $60$ eV, respectively.
The fit is significantly improved and becomes acceptable 
($\chi^2/{\rm d.o.f.}= 313/361$) with this model 
{\tt phabs*(xsabs*simpl*bhspec+gauss+gauss)} where the solar abundances are
assumed for both the ionized and neutral absorbers.  For the ionized absorber, 
the resulting ionization parameter is $\log \xi =  1.7 \pm 0.1$ and the 
column density is $(1.4^{+0.4}_{-0.3}) \times 10^{22}$ cm$^{-2}$. 
However, the equivalent width of the iron-K absorption line estimated from this model 
is somewhat smaller than that actually seen in the {\it Swift}/XRT
spectrum. We therefore add another ionized absorption component and fit
the spectrum with the model expressed as {\tt
phabs*(xsabs*xsabs*simpl*bhspec+gauss+gauss)} (Model 3 in Table~\ref{tab_xrt}).  
This model excellently describes the overall spectrum and further decreases 
the reduced chi-squared value ($\chi^2/{\rm d.o.f.} = 294/359$). The resulting 
model is plotted in Figure~\ref{fig_xrt} and the best-fit parameters are given 
in Table~\ref{tab_xrt}.  The ionization parameter and column density are
$\log \xi = 2.86^{+0.52}_{-0.18}$ and $5.1^{+3.8}_{-2.8} \times
10^{22}$ cm$^{-2}$ for one ionized absorber responsible for the iron
K$\alpha$ absorption line, and $\log \xi = 1.2 \pm 0.2$ and
$1.0^{+0.4}_{-0.3} \times 10^{22}$ cm$^{-2}$ for the other. The spin
parameter is estimated as $a = 0.46 \pm 0.06$.

For direct comparison of the disk emission between the low/hard state
({\it Suzaku}) and the high/soft state ({\it Swift}), we replace {\tt bhspec} in the
final model with {\tt diskbb} ({\tt phabs*(xsabs*xsabs*simpl*diskbb+gauss+gauss)}; 
Model 4 in Table~\ref{tab_xrt}) and fit the {\it Swift}/XRT spectrum. The 
fit is again acceptable ($\chi^2/{\rm d.o.f.} = 292/359$), and the inner disk 
temperature and normalization of {\tt diskbb} are estimated to be 
$0.88^{+0.03}_{-0.04}$ keV and $139^{+26}_{-45}$, respectively. 
The normalization is about 35 
times smaller than that of the direct MCD component obtained from the
{\it Suzaku} best-fit model. This indicates that the inner disk radius is
smaller during the {\it Swift} observation in the high/soft state than during
the {\it Suzaku} observation in the low/hard state.  These {\it Swift}/XRT results
are summarized in Table~\ref{tab_xrt} and Figure~\ref{fig_xrt}.

\section{Near-infrared Observations and Results}

Photometric observations of MAXI~J1305$-$704 in the $J$ (1.25$\mu$m),
$H$ (1.63$\mu$m), and $K_{\rm s}$ (2.14$\mu$m) bands were carried out on
6 nights by using the SIRIUS camera \citep{nag03} on the 1.4m {\it IRSF}
telescope at the South African Astronomical Observatory (SAAO). The 
first 3 nights (2012 April 27, 28, and 29) were about 20 days after 
the beginning of outburst and the source was stayed in the high/soft 
state, while it was in the low/hard state on the last 3 nights 
(2012 July 22, 23, and 24). The July observations with {\it IRSF}/SIRIUS 
were made only 1 day after the end of {\it Suzaku} X-ray observation. 
The typical seeing in full width at half maximum was $\approx$ 1.5--2.0'' 
(3.5--4.5 pixels) in the $J$ band. The observation log is given 
in Table~\ref{tab_irsf}.

\begin{table*}
\begin{center}
\caption{Log of {\it IRSF} observations.\label{tab_irsf}}
\begin{tabular}{cccccc}
\tableline
Date & Number of  & Integration time  & 
\multicolumn{3}{c}{Magnitude\tablenotemark{b,c}} \\
(2012) & observations\tablenotemark{a} & in each frame (sec) & $J$ & $H$ & $K_{\rm S}$ \\ 
\tableline
Apr. 27 & 1  & 30 & $15.95 \pm 0.53$ & $15.69 \pm 0.57$ & $14.99 \pm 0.59$ \\
Apr. 28 & 1  & 30 & $15.84 \pm 0.51$ & $15.49 \pm 0.49$ & $15.32 \pm 0.59$ \\
Apr. 29 & 54 & 30 & $15.74 \pm 0.47$ & $15.43 \pm 0.47$ & $15.19 \pm 0.47$ \\
Jul. 22 & 1  & 15 & $16.63 \pm 0.10$ & $16.20 \pm 0.05$ & $15.86 \pm 0.09$ \\
Jul. 23 & 1  & 15 & $16.46 \pm 0.04$ & $16.08 \pm 0.04$ & $15.88 \pm 0.08$ \\
Jul. 24 & 1  & 15 & $16.66 \pm 0.05$ & $16.24 \pm 0.03$ & $16.03 \pm 0.08$ \\
\tableline
\end{tabular}
\tablenotetext{a}{10 and 15 dithered frames are combined for each observation at first and last 
three nights, respectively.}
\tablenotetext{b}{All the object frames are added for each night separately to measure the magnitudes.}
\tablenotetext{c}{Systematic errors due to the installation of the polarizer (3\% of the observed 
magnitude at the maximum) are included in the data of the April observations.}
\end{center}
\end{table*}

\begin{figure*}[tbp]
\plottwo{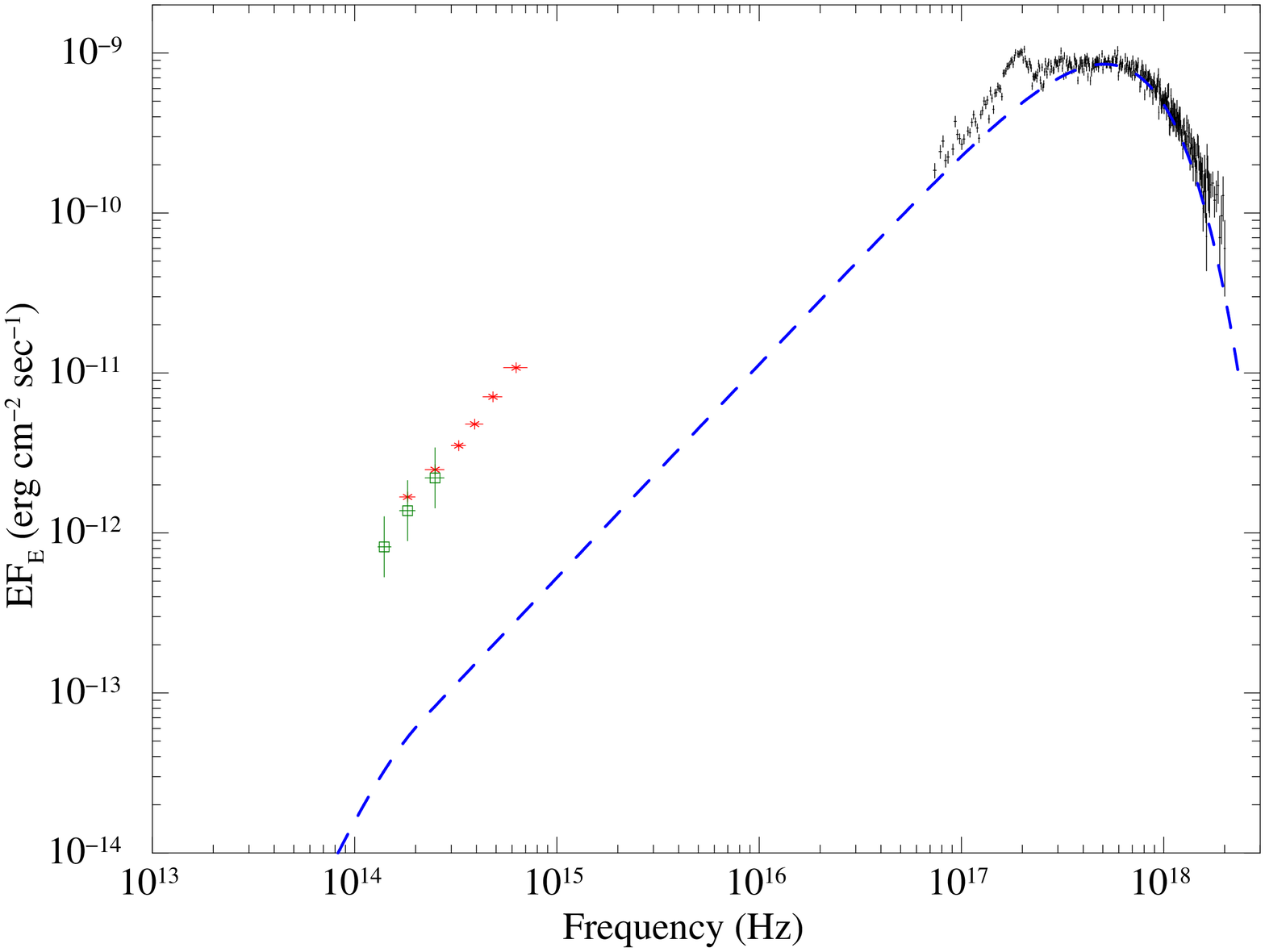}{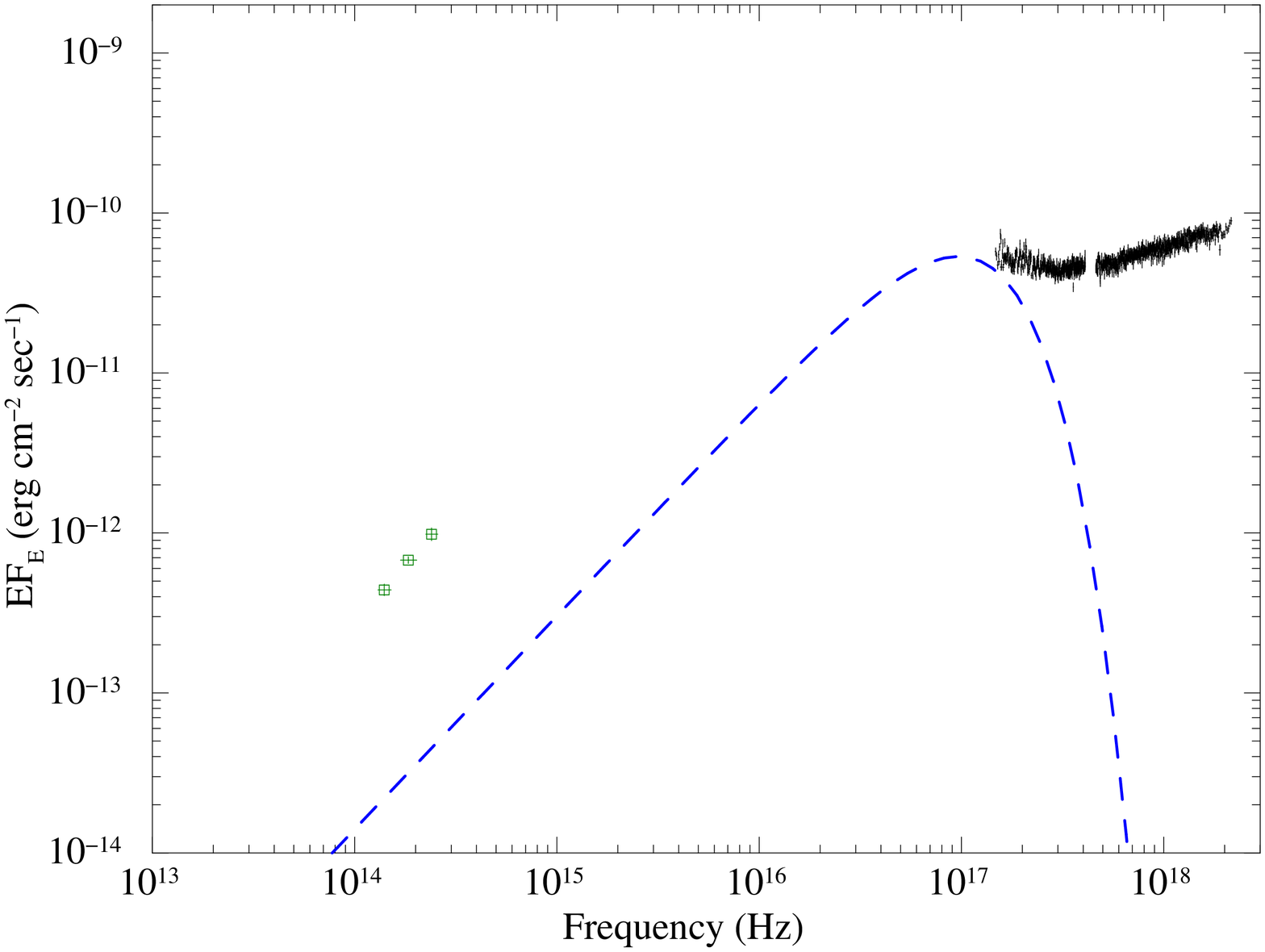}
\caption{The spectral energy distributions of MAXI~J1305$-$704 in the high/soft state 
and low/hard state are plotted in the left and right panels, respectively. 
The {\it IRSF} fluxes in the $J$ (1.25 $\mu$m), $H$(1.63 $\mu$m), and $K_S$ 
(2.14 $\mu$m) bands obtained on 2012 April 29 (left) and July 22 (right), 
which are corrected for interstellar extinction (green, open square). 
The black points are the {\it Swift}/XRT (left) and {\it Suzaku} (right) 
spectra, corrected for neutral and ionized absorptions. Blue dashed lines 
shows the intrinsic disk spectra including the Comptonized photons, 
where the outer disk radius is assumed to be the Roche lobe size. 
The optical and near infrared fluxes obtained from the GROND observation 
on April 11 are also shown in the left panel (red cross). 
\label{fig_irsf}}
\end{figure*}

We performed the standard data reduction (i.e., dark subtraction,
flat-fielding, sky subtraction, and combining dithered images) with {\it IRSF}
pipeline software on IRAF version 2.16 (the Image Reduction and Analysis 
Facility, distributed by the National Optical Astronomy Observatory). 
We combined all the object frames obtained in one night to maximize the 
signal-to-noise ratio. 
We found the most probable near infrared counterpart of MAXI
J1305$-$704 at R.A.\ $= 13^{\rm h}06^{\rm m}55^{\rm s}.3 \pm 0^{\rm s}.1$ 
and Dec.\ $=-70^\circ27'05''.1 \pm 0''.1$ (J2000), 
which is located in the {\it Swift}/XRT 90\% error circle \citep{ken12} 
and is consistent with the position of the optical/near-infrared 
counterpart discovered on April 11 by \citet{gre12} with the GROND 
instrument mounted on the 2.2m telescope
in the MPI/ESO La Silla observatory. The {\it IRSF}/SIRIUS position is
also consistent with those estimated in the {\it Swift}/UVOT
(Ultraviolet/Optical Telescope) and {\it Chandra} HETGS observations
\citep{gre12,mil12b} performed on April 10 and 29, respectively. The
magnitudes in the three bands on each night are listed in
Table~\ref{tab_irsf}. These were obtained by performing aperture
photometry calibrated with the 2MASS 
\citep[2 Micron All-SkySurvey;][]{skr06} photometric data of the stars 
in the field of view.

Figure~\ref{fig_irsf} shows the {\it IRSF} fluxes on April 29 in 
the high/soft state (left panel) and on July 22 in the low/hard state 
(right panel). These fluxes were corrected for Galactic extinction. 
Considering the {\it Suzaku} and {\it Swift} results, we assumed the 
hydrogen column density of interstellar absorption as 
$N_{\rm H} = 1 \times 10^{21}$ cm$^{-2}$ and derived the extinction in 
each band as $A_{\rm J} = 0.15$, $A_{\rm H} =0.09$, and $A_{\rm K} = 0.06$ 
by combining the conversion factors given by \citet{pre95}
and \citet{rie85}. 

In Figure~\ref{fig_irsf}, the quasi-simultaneous X-ray data obtained in
the {\it Swift} and {\it Suzaku} observations are plotted in the left and 
right panel, respectively. The best-fit intrinsic disk components are
separately shown. The X-ray spectra are corrected for both neutral and
ionized absorptions. 
The GROND data in the optical and near infrared bands 
\citep[$g'$, $i'$, $r'$, $z'$, $J$, and $H$;][]{gre12} are also 
plotted in Fig.~\ref{fig_irsf} together with our {\it IRSF} ones 
obtained in the high/soft state. 
As noticed from the figure, these fluxes in the high/soft and 
low/hard states are $\approx 10$ times larger than those of the 
intrinsic disk components estimated from the X-ray data. 
The flux levels in the optical and near infrared bands were decreased 
by $\approx$ 50\% from the high/soft state to the low/hard state. 
These results suggest that in addition to the direct disk emission 
and the constant black body radiation from the companion star, another 
component (probably irradiation in the outer disk region) significantly 
contributes to the optical and near infrared fluxes (see also Section 5.1).
The {\it IRSF} fluxes on July 24, which are the weakest ones in the 6 
nights, correspond to the absolute magnitudes of 2.8, 2.4, and 2.1 in 
the $J$, $H$, and $K_{\rm S}$ bands (where the distance of MAXI 
J1305$-$704 is assumed as 6 kpc), respectively. If the companion is 
a main-sequence star, these magnitudes indicate that it is a late 
F-type or smaller mass star \citep{wai92}.

\section{Discussion}

\subsection{Implications for the System Parameters}

It is likely that the compact object of MAXI J1305$-$704 is a black hole
because the behavior of spectral evolution in the outburst is quite
similar to those of typical BHXBs \citep{mor13}. However, no constraint
has been obtained so far on the black hole mass of this source, as well
as its distance and the mass of the companion star. Here we summarize
what we find about these system parameters from the {\it Suzaku} and 
{\it Swift} results.

The power spectrum obtained with the XIS light curve shows very weak 
intrinsic variability with a fractional rms$^2$ of $\sim 10^{-3}$ 
Hz$^{-1}$ from $1 \times 10^{-3}$ Hz to $5\times 10^{-2}$ Hz. This result 
suggests that the low-frequency break of the band limited noise is 
located above the frequency range. 
Normally, BHXBs have an order of magnitude stronger power and the break
frequency is much lower when they are in the low/hard state, although weaker 
variability is sometimes observed from low mass black holes 
like GRO~J1655$-$40 (5--7 $M_\sun$, \citealt{rem99}) in that state. This might 
suggest that MAXI~J1305$-$704 also have a relatively small mass black hole. 
However, even with a low mass black hole it is difficult to explain the lack 
of the variability power for such a very hard spectrum with a photon index 
of $\approx 1.6$.

Since the source shows dips but no eclipse, its inclination angle $i$
is estimated to be $\approx 60^\circ$--$75^\circ$ \citep{fra87}. The
dips seen in MAXI J1305$-$704 are deeper and more periodic than those
in GRO J1655$-$40, whose inclination angle is $69^\circ.50 \pm
0^\circ.08$ \citep{oro97}. This suggests that MAXI J1305$-$704 has a
larger inclination angle than GRO J1655$-$40, likely $\approx
75^\circ$, and that the complex dips are originated in absorbing
structures with small scale heights above the disk crossing the line
of sight. From the {\it Suzaku} XIS light curve, the dip 
interval is estimated as $9.74 \pm 0.04$, which likely correspond 
to the orbital period of MAXI~J1305$-$704. 
We derive the binary size as $\approx 3 \times 10^6 M_{{\rm tot} 4}^{1/3}$ km 
from the Kepler's third law, where $M_{{\rm tot} 4}$ represents the total 
mass of companion star and black hole in the unit of $4 M_{\sun}$.
Combining the Kepler's third law and the relation between the radius 
and mass of the Roche lobe in a semi-detached binary system 
\citep[Equation~4. in][]{pat71}, 
we have
\begin{equation}
\rho_{\rm c} = 30.375 \, \frac{\pi}{G P^2} \;  (0 < M{\rm c}/M_{\rm BH} < 0.8),\label{eq2}
\end{equation}
where $\rho_{\rm c}$, $M_{\rm c}$, and $P$ represent the averaged density 
and mass of the companion star, which fills its Roche lobe, and the 
orbital period. From this equation, we derive the averaged density of the 
companion star as $\approx 1.2$ g cm$^{-3}$, which is smaller than 
that of the Sun ($\approx 1.4$ g cm$^{-3}$). If the companion is 
a main sequence star, it has a slightly larger mass than the Sun. 
This is consistent with the near-infrared absolute magnitudes observed 
with {\it IRSF} in the low/hard state. However, an upper limit of the 
stellar radius is imposed by the inclination angle and binary size, 
$\approx 7 \times 10^5$ $(\cos i/\cos 75^\circ)$ $M_{{\rm tot} 4}^{1/3}$ km, 
or $\approx 1$ $(\cos i/\cos 75^\circ)$ $M_{{\rm tot} 4}^{1/3}$ $R_\sun$, 
by considering that the source has no eclipses. This radius and the 
averaged density of the Roche lobe give a somewhat smaller mass of 
the companion star than that of the Sun, 
$< 0.9$ $(\cos i/\cos 75^\circ)^3$ $M_{{\rm tot} 4}$ $M_\sun$, although 
this limit strongly depends on the assumed inclination angle and the 
total mass of the binary system. Thus, it is also possible that the 
companion is an evolved star with a mass of $\lesssim 1 M_\sun$, 
instead of an earlier-type main sequence star than the Sun.

As presented in Fig.~\ref{fig_irsf}, the near infrared and optical 
fluxes of MAXI J1305$-$704 in the high/soft state is $\approx$ 10 times 
higher than the flux level of the multicolor disk component, suggesting 
that the fluxes are dominated by other components, likely reprocessed 
emission from the irradiated outer disk and the black body emission 
from the companion star. To estimate the contributions of these two 
components to the optical and near infrared spectral energy distribution 
(SED), we fit the {\it Swift}/XRT (X-ray), GROND (optical and near 
infrared), and {\it IRSF} (near infrared) data in the high/soft state using 
the {\tt diskir} \citep{gie08,gie09} plus {\tt bbodyrad} model, which represent 
the direct and reprocessed emission from the disk and the black body component 
from the companion star, respectively. The {\tt diskir} model calculates the 
total spectrum of the disk emission and its Comptonization including the 
reprocessed emission from the irradiated outer disk, by using the inner 
disk temperature ($kT_{\rm in}$), photon index and electron temperature 
of the Comptonized component ($\Gamma$ and $kT_{\rm e}$, respectively), 
the ratio of the luminosity of the Compton tail to disk luminosity 
($L_{\rm C}/L_{\rm d}$), the fraction of luminosity of the Comptonized 
component that is thermalized in the inner disk ($f_{\rm in}$), the fraction 
of bolometric flux that illuminates the outer disk ($f_{\rm out}$), 
the radius of the Compton illuminated disk ($r_{\rm irr}$), and the outer disk 
radius ($R_{\rm out}$). The {\tt bbodyrad} model produces a black body 
spectrum from a temperature ($kT_{\rm BB}$) and a normalization ($K_{\rm BB}$), 
which is related to the source radius $R_{\rm BB}$ (km) and distance 
through $K_{\rm BB} = (R_{\rm BB}/D_{10})^2$, where $D_{10}$ is the distance 
in units of 10 kpc.

We replace {\tt diskbb} of the best-fit results of Model 4 in the 
{\it Swift}/XRT fit (Section~3.4) with {\tt diskir} and add {\tt bbodyrad} 
to fit the multiwavelength SED. Here we set $f_{\rm in} = 0.1$ and 
$r_{\rm irr} = 1.1 R_{\rm in}$ following \citet{gie09}. The photon index 
of the Compton tail and the inner disk temperature are fixed at the same 
values of Model 4, $\Gamma = 2.2$ and $kT_{\rm in} = 0.88$ keV. The electron 
temperature of the Compton component is set to 300 keV in order not to have 
an exponential cut-off in the energy range of the {\it Swift}/XRT. 
After some traials of the spectral fit, we find that the black body 
component favors a small temperature and a large normalization.
Considering the maximum radius constrained from the absence of eclipse, we vary 
the normalization of the {\tt bbodyrad} component within 
$R_{\rm BB} < 1.0 D_6$ $R_\sun$.

\begin{figure}[tbp]
\plotone{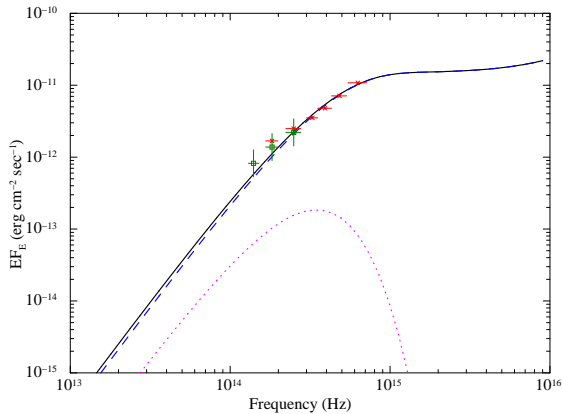}
\caption{The best-fit {\tt diskir+bbodyrad} spectrum in the optical and near infrared 
bands. The contributions of {\tt diskir} (blue, dashed) and black body (pink, dotted) 
components are separately displayed. The GROND fluxes obtained on April 11 (red, cross) and 
{\it IRSF} ones on April 29 (green, open square) are overplotted. \label{fig_diskir}}
\end{figure}

The best-fit parameters of the {\tt diskir} and {\tt bbodyrad} 
models are listed in Table~\ref{tab_diskir}. The optical and near 
infrared SED in the high/soft state is reproduced reasonably well 
with the dominant irradiated disk component and the weak black body 
component from the companion star (Figure~\ref{fig_diskir}). The 
irradiation fraction ($f_{\rm out} \approx 4 \times 10^{-3}$) is 
$\approx$ 10 times smaller than those of GX 339$-$4 and 
GRS 1915$+$105 during the high/soft state \citep{rah10, rah12}, 
but $\approx$ 5 times larger than those of XTE J1817$-$330 
in that state \citep{gie09}. The radius of the companion star is 
estimated as $R_{\rm BB} > 0.8 D_6$ $R_\sun$ and the temperature is 
found to be smaller than that of the Sun (3000 K $< T_{\rm BB} <$ 
5600 K). These results and the averaged density 
estimated from Equation~\ref{eq2} suggest that the companion is 
likely an evolved star at an early stage that has a smaller mass 
than $\approx 1M_\sun$.

\begin{table}
\begin{center}
\caption{Best-fit results from the simultaneous fit of the X-ray, 
optical, and near infrared data of the {\it Swift}/XRT, GROND, and {\it IRSF}.
\label{tab_diskir}}
\begin{tabular}{lcc}
\tableline\tableline
Component & Parameter & value \\
\tableline
diskir & $kT_{\rm in}$ (keV) & $0.88$ (fix) \\
 & $\Gamma$ & 2.2 (fix) \\
 & $kT_{\rm e}$ (keV) & 300 (fix) \\
 & $L_{\rm C}/L_{\rm d}$& $0.10 \pm 0.02$ \\
 & f$_{\rm in}$ & 0.1 (fix) \\
 & $r_{\rm irr}$ ($R_{\rm in}$) & 1.1 (fix) \\ 
 & $f_{\rm out}$ & $4.4^{+0.05}_{-0.04} \times 10^{-3}$ \\
 & $\log_{10} (R_{\rm out}/R_{\rm in})$ & $5.15^{+0.4}_{-0.5}$ \\ 
 & norm  & $137 \pm 1$ \\
bbodyrad & $T_{\rm BB}$ (K) & $4200^{+1400}_{-1200}$ \\
 & $R_{\rm BB}$ ($R_\sun$/6 kpc) & $1.0^{+0,\; {\rm pegged}}_{-0.2}$ \\
\tableline
 $\chi^2/{\rm d.o.f}$ & & $415/377$ \\
\tableline
\end{tabular}
\tablecomments{The model is 
{\tt phabs*(xsabs*xsabs*diskir$+$bbodyrad$+$\\gauss$+$gauss)}. The parameters of 
neutral and ionized absorptions ({\tt phabs} and {\tt xsabs}, respectively) and two 
Gaussians are set to the best-fit values of Model 4 in the {\it Swift}/XRT spectral fit 
(Table~\ref{tab_xrt}).}
\end{center}
\end{table}

We find that the non-dip spectra in the {\it Swift} and {\it Suzaku} 
observations have comparable bolometric luminosities within a factor of 3. 
This suggests that they are likely to be about only a few times more and less 
than that in the soft-to-hard transition, $\approx 0.02 L_{\rm Edd}$, respectively,
by considering the results of \citet{mcc03}. However, the inner disk
temperature of the {\tt diskbb} component obtained from the {\it Swift}
spectrum ($\approx 1.0$ keV) is much higher than what we expect from a
faint high/soft state for normal BHXBs. This suggests that the disk
spectrum is significantly modified by the Doppler effects due to a very
high inclination angle, probably $\approx 75^\circ$. Indeed, assuming a
black hole mass of $3 M_{\sun}$, bolometric luminosity of $0.05 L_{\rm Edd}$, 
and inclination angle of $75^\circ$, we successfully reproduce
the {\it Swift} spectrum with a relativistic disk emission model ({\tt
bhspec}), and obtain a moderate spin ($a = 0.46 \pm 0.06$) and a
distance of $6.3^{+0.4}_{-0.3}$ kpc calculated from the normalization 
($2.5^{+0.3}_{-0.2}$). 
We note that there is strong coupling between the spin,
luminosity, and black hole mass in the {\tt bhspec} fit and a smaller
spin parameter is obtained when a larger luminosity and/or a lower black
hole mass is assumed. For instance, in the cases of $L_{\rm X} = 0.1
L_{\rm Edd}$ and $0.01 L_{\rm Edd}$, we have $a< 0.16$ and $0.90 \pm 0.02$, 
while $M_{\rm BH}= 5 M_\sun$ and $10 M_\sun$ give $a=0.63^{+0.05}_{-0.04}$ and 
$0.80 \pm 0.03$, respectively.

\subsection{Dipping Behavior}

The {\it Suzaku} observation revealed that MAXI J1305$-$704 has two separate
periodic dips with different column densities, ionization parameters,
and covering fractions of the absorbers. We find that these dips have
the same recurrence period of $9.74 \pm 0.04$ hours and the harder dip
(deep dip) is followed by the softer one (shallow dip) in 6.38
hours. Such strong, softer ``secondary dips'' are occasionally seen in
dipping X-ray binaries (a few neutron star binaries like XB 1916$-$053;
\citealt{sma92}). Dips are generally interpreted as the absorption by
the ``bulge'' formed in a region where the accretion stream from the
companion star impacts the outer boundary of the disk, but what is the
origin of secondary dip? It may be the result that the accretion stream
hits the disk again and splashes at its circularization radius, which is
smaller than the outer disk radius \citep{fra87, fra92}. \citet{fra87}
suggest that the stream causes ionization instabilities at the second
impact and creates patchy cold clouds within hot medium. This is
consistent with the behavior of the shallow dip, in which significant
time variabilities can be seen in the XIS light curve. The shallow dip
occurs at the orbital phase of $\approx$ 0.64 (if the start of the deep
dip is assumed to be phase 0), which is also consistent with the picture
of \citet{fra87}.

The properties of time variabilities and absorption profiles in the
dipping spectra provide us with key information on the two dips. 
The shallow dip exhibits significant fast variabilities on the timescale of 
a few minutes in the XIS light curve, suggesting that the absorber is not 
a single continuous structure, but composed of blobs. 
If the absorber of the shallow dip is created at the 
circularization radius of $10^6$ km and rotates with the
Keplerian velocity, one minute corresponds to a typical blob size
of $\approx 4 \times10^4 M_3^{1/2}$ km, where $M_3$ is defined as 
$M_{\rm BH}/3 M_\sun$. 
The covering fraction $0.72^{+0.03}_{-0.04}$ in the shallow dip can be
understood as the filling factor of blobs in the dipping zone. 
Interestingly, similar short-time variability was also found in the 
neutron star XB 1254$-$690 \citep{dia09}, suggesting that shallow 
dips in dipping X-ray binaries may generally consist of clumps.
By contrast, in the deep dip, short-time variabilities are not significant
and the covering fraction is larger than 90\%, suggesting
that the absorber of the deep dip has more continuous structure or is
filled with much smaller blobs than those of the shallow dip.

We have shown that both non-dip and dip spectra of MAXI~J1305$-$704
obtained with {\it Suzaku} are successfully modeled by ionized absorbers with
different column densities and ionization parameters. The dip spectra
have an order of magnitude larger hydrogen column densities and 
smaller ionization parameters than those of the non-dip spectrum. These
results are very similar to those reported by \citet{boi05} and
\citet{dia06} for neutron star low mass X-ray binaries. Thus, it may be
a general picture in low-mass X-ray binaries that dips are created by
ionized absorbers with much larger column densities and in lower ionization
states than those observed in non-dip spectra.

\subsection{Structure of Accretion Disk and Comptonized Corona}

The {\it Suzaku} non-dip spectrum of MAXI~J1305$-$704 is approximated by a
power-law extending up to 130 keV with a photon index of $\approx 1.6$. 
This hard spectrum indicates that the source was in the low/hard
state. By more detailed modeling, we find that the spectrum can be
described with a general model in the low/hard state of BHXBs 
\citep[e.g.,][]{gie97}, a multicolor disk and its Comptonization with a 
reflection component from the disk. Although recent {\it Suzaku} studies 
on other BHXBs report that two Comptonization components with different 
optical depths are needed to reproduce their spectra in the low/hard 
state \citep{tak08,mak08,shi11,yam13}, our data do not require the second
component. The reason is unclear, but it may be because the large inclination 
of MAXI~J1305$-$704 makes it difficult to detect the softer component 
(i.e., with a small optical depth) than the other systems, although it 
may be partially due to the poor statistics of our data. 
The smooth spectral profile of MAXI J1305$-$704 without any complex 
Comptonized component required, as well as the relatively weak 
time variability below $5 \times 10^{-2}$ Hz (see Section 5.1) are 
unusual and interesting properties for a BHXB in the low/hard state.

We obtained a much larger normalization ($6.0^{+3.4}_{-2.4} \times 10^3$) 
and smaller temperature ($0.168^{+0.008}_{-0.006}$ keV) of the
direct MCD component than those obtained with the {\it Swift}/XRT spectrum in
the high/soft state. Assuming that the Comptonizated
corona is isotropic and that the total number of photons from the disk
is conserved after reprocessed by Comptonization, we obtain the
following equation \citep{kub04};
\begin{eqnarray}
F^p_{\rm disk}+F^p_{\rm thc}2 \cos i &= 0.0165 \left[ \frac{r_{\rm in}^2\cos i}{(D/10\mbox{ kpc})^2}\right] 
\left( \frac{T_{\rm in}}{1\mbox{ keV}} \right)^3 \nonumber \\
& \mbox{ photons } {\rm s}^{-1} \mbox{ }{\rm cm}^{-2}, \label{eq1}
\end{eqnarray}
where $F^p_{\rm disk}$ and $F^p_{\rm thc}$ are the 0.01--100 keV photon 
flux from the disk and thermal Comptonized component, respectively. 
We estimate the flux of the {\tt nthcomp} component as $0.180$ photons 
cm$^{-2}$ sec$^{-1}$ and that of diskbb component as $0.394$ 
photons cm$^{-2}$ sec$^{-1}$. Using Equation~\ref{eq1}, we estimate the 
innermost disk radius of $r_{\rm in}=93^{+7}_{-5} D_6 (\cos i/\cos 75^\circ)^{-1/2}$ km, 
(where $D_6$ is the distance in unit of 6 kpc). The actual radius is derived 
to be $R_{\rm in} = 111^{+8}_{-6} D_6 (\cos i/\cos 75^\circ)^{-1/2}$ km, by 
multiplying 1.19, a correction factor of the boundary condition and spectral 
hardening \citep{kub98}. 

We compare the inner disk radius in the {\it Suzaku} and {\it Swift} observations,
using the intrinsic flux of the MCD component. Although the absolute
radius obtained from the {\it Swift} result may be affected by the strong
beaming effects, we are able to discuss the relative difference of the
radius between the two epochs. Using the {\tt diskbb} normalization
obtained with the XRT spectrum ($139^{+26}_{-45}$) and multiplying the
correction factor 1.19 \citep{kub98}, we derive the inner disk radius as
$16.5^{+1.5}_{-2.9} D_6 (\cos i/\cos 75^\circ)^{-1/2}$ km.  Thus, the
inner radius obtained from the best-fit model of {\it Suzaku} data, 
$111^{+8}_{-6} D_6 (\cos i/\cos 75^\circ)^{-1/2}$ km is 5.8--8.8 times larger than
that from the XRT result. Thus, we robustly conclude that the inner
radius increased in the {\it Suzaku} observation, giving strong evidence for
disk truncation in the low/hard state.

We find significant broad emission-line like residuals at $\approx$ 0.7
keV and $\approx$ 1.2 keV in the {\it Swift}/XRT spectrum, which cannot 
be reduced by multiple ionized absorptions, partial covering, emission
components from the absorber itself.
Also, these residuals are not completely explained by changing the 
elemental abundances in the neutral and ionized absorbers. 
It is difficult to know what makes these structures. Because they 
are not seen in the Suzaku spectra in the low/hard state, their origin 
would be associated with the geometry of the accretion disk and/or 
the ionized absorbers in the high/soft state.

Similar features at $\approx$ 0.7 keV were reported in several 
ultracompact low mass X-ray binaries 
\citep[UCXBs; e.g.,][]{jue01}, which have very short orbital 
periods (less than about 80 minutes). \citet{mad10} and \citet{mad11} 
recently suggested that the structures seen in UCXBs are the 
relativistically broadened OVIII Ly$\alpha$ line created by 
reflection on the disk in the vicinity of the accretor 
with an ionization parameter of $\log \xi \approx 2.3$.
Likewise, the residuals in MAXI~J1305$-$704 might be originated 
from reflection on ionized accretion disk in the strong gravitational 
field created by the central black hole. 
Indeed, we find a relativistic emission line model {\tt laor} instead 
of Gaussian can also give an acceptable fit 
($\chi^2/{\rm d.o.f.} = 297/360$) by assuming an inclination of 
$75^\circ$, inner radius of $4 R_{\rm g}$ (corresponding to $a \approx 0.5$), 
outer radius of $400 R_{\rm g}$, and emissivity index ($\beta$). 
The resulting line energy and equivalent width of the {\tt laor} model 
are $0.66 \pm 0.01$ keV and $194$ eV for the lower energy feature 
and $1.03 \pm 0.02$ keV and $83$ eV for the higher energy one (here 
we vary the emissivity index within $2 \leq \beta \leq 3$ and obtain 
the best fit value of $\beta = 2.4 \pm 0.1$). These line energies 
are consistent with the K$\alpha$ line from H-like oxygen ions and 
L-lines from ionized iron ions.
The photons to illuminate the disk would be produced from the 
Comptonizing corona, or from the disk itself, whose emission could be
partially incident on the disk because of gravitational light bending.
If these photons could strongly illuminate the disk, huge emission 
lines might arise through a temperature inversion region of the 
irradiated disk atmosphere. 
The large equivalent width of the emission lines with respect 
to the continuum emission would be expected for a very high inclination 
source because of a large optical depth of the disk atmosphere. 
However, we have no model to accurately evaluate these effects 
at present. Radiative transfer calculation including all these 
possible effects is left for future studies.

\subsection{Ionized absorbers}

We find that not only the {\it Swift}/XRT spectrum in the high/soft state but
also the {\it Suzaku} spectrum in the low/hard state exhibit ionized
absorption features. Blue shifts are not significantly detected with
upper limits of 2300--5800 km s$^{-1}$. The absorber of the {\it Suzaku} non-dip
spectrum has a column density of $N_{\rm H} = 6.1^{+1.0}_{-0.9} \times
10^{21}$ cm$^{-2}$ and an ionization parameter of $\log \xi =
2.19 \pm 0.04$, while the {\it Swift}/XRT spectrum requires two ionized
absorbers with different parameters, $N_{\rm H} = 5.1^{+3.8}_{-2.8} \times
10^{22}$ cm$^{-2}$ and $\log \xi = 2.86^{+0.52}_{-0.18}$ for one, $N_{\rm H} = 
(1.0^{+0.4}_{-0.3}) \times 10^{22}$ cm$^{-2}$ and $\log \xi = 1.2 \pm 0.2$
for the other. As discussed in Section 5.2, the dip spectra of {\it Suzaku} 
are also described with ionized absorbers but with much larger column
densities and lower ionization parameters, $N_{\rm H} = (1.44 \pm 0.06)
\times 10^{23}$ cm$^{-2}$ and $\log \xi = 1.90 \pm 0.07$ for the deep 
dip, and $N_{\rm H} = 6.6^{+0.5}_{-0.4} \times 10^{22}$ cm$^{-2}$ and $\log \xi =
1.79 \pm 0.07$ for the shallow dip. 

To investigate the origin of the ionized absorbers, we estimate their distances 
from the X-ray source ($R$) from the definition of the ionization parameter, 
\begin{equation}
\xi = \frac{L_{\rm X}}{n_{\rm H} R^2}  = \frac{L_{\rm X}}{N_{\rm H} R} \frac{\Delta R}{R}, 
\end{equation}
where $n_{\rm H}$ and $\Delta R$ represent the hydrogen number-density and the length
of the absorber, respectively. Here $L_{\rm X}$ is the incident luminosity, whose 
energy range is 1--1000 Ry in the definition of XSTAR. 
$L_{\rm X}$ is estimated as $1.3 \times 10^{36} D_6^2$ erg sec$^{-1}$ for the {\it Suzaku} 
observation, and $8.1 \times 10^{36} D_6^2$ erg sec$^{-1}$ for the {\it Swift} observation.
Assuming $\Delta R/R = 1$, we obtain $R \approx 1.4 \times 10^7$ km for the absorber 
seen in the {\it Suzaku} non-dip spectrum. For the {\it Swift}/XRT data, we estimate 
$R \approx 2.2 \times 10^{6}$ km for the absorber with a higher ionization state 
and $R \approx 5.1 \times 10^{8}$ km for that with a lower ionization state. 
All these radii are more than an order of magnitude larger than predicted values for 
thermally driven disk winds in BHXBs ($R \sim 10^5$ km; e.g., \citealt{beg83}; 
\citealt{woo96}; also see below). Furthermore, 
they are comparable to or even larger than the binary size (see Section 5.1). 
Hence, if the ionized absorbers of MAXI~J1305$-$704 are located on the disk, 
$\Delta R/R$ should be much less than 1. This suggests that the
absorbers are originated from rather compact structures and do not
largely extend in the radial direction. The absorbers in the deep and
shallow dips are calculated to be $R \approx 1.1 \times 10^6$ km, and $R
\approx 3.2 \times 10^6$ km, respectively, which are comparable to those
seen in the {\it Suzaku} and {\it Swift}/XRT non-dip spectra.

Table~\ref{tab_wind} summarizes the physical properties of ionized absorbers
reported in previous studies of black hole or neutron star X-ray
binaries. The ionization parameter, column density, Doppler velocity,
and luminosity taken from the literature are listed.
We then calculate the apparent distance $R$ by assuming 
$\Delta R/R = 1$ for each set of parameters. Although there is considerable 
variety in each parameter, almost all of the absorbers in the BHXBs are found 
to be outflowing with a velocity of 100--1000 km sec$^{-1}$. Typically, they 
are located at $R \sim 10^4$--$10^5$ km when $\Delta R \sim R$ is
assumed. They are interpreted as thermally 
\citep[e.g.,][]{kub07}, radiatively \citep{kot00}, or magnetically 
driven disk winds \citep{mil08}.
By contrast, ionized absorbers in neutron star low mass X-ray binaries 
do not often exhibit significant blue shifts, except for some sources 
like GX 13+1 \citep{ued04} and Cir X-1 \citep{sch08}, which are known 
to have powerful outflows. This implies that the photoionized plasma 
on neutron star dippers remain gravitationally bound to the system as 
disk atmosphere and is not outflowing due to a small system size and a 
low luminosity of the central X-ray source \citep[see][]{dia12}.

\begin{figure}
\plotone{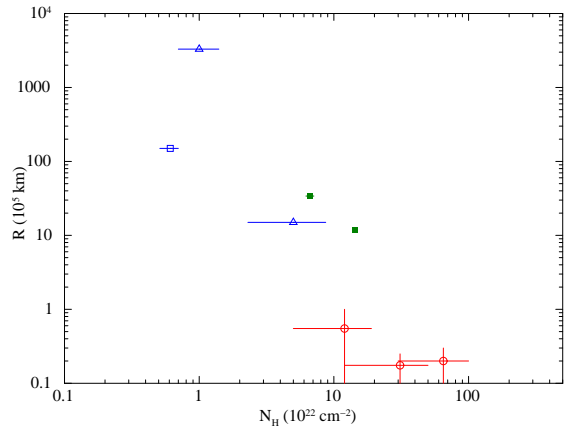}
\caption{$N_{\rm H}$ vs. $R$ for ionized absorbers in persistent spectra of 
MAXI J1305$-$704 (blue) in the high/soft state (open triangle) and 
low/hard state (open square) and other BHXBs that exhibit dips (red circle), 
except for GRS 1915$+$105, taken from Table~\ref{tab_wind}. 
Only the absorbers whose turbulent velocities are less than 1000 km 
sec$^{-1}$ are plotted.
The absorbers of MAXI~J1305$-$704 in the dipping periods are 
also plotted (green, filled square).\label{fig_xsabs}}
\end{figure}

From the comparison, we find that MAXI~J1305$-$704 has somewhat lower
ionization parameters ($\log \xi < 3$) and consequently larger $R$ than those 
of typical disk winds observed in other BHXBs. By contrast, absorbers in 
neutron star dippers sometimes have similar $R$ values (see the table of 
XB 1916$-$053 and EXO 0748$-$676) and exhibit complex and deep dips like 
MAXI J1305$-$704. Although the BHXB GRO~J1655$-$40 shows a similar 
value of $R$ ($4 \times 10^6$ km 
and $1.3 \times 10^7$ km) based on the results by \citet{dia07}, the turbulent 
velocity ($v_{\rm turb}$) adopted there is much higher than that assumed for 
MAXI~J1305$-$704 in our paper. We have to note that a larger $v_{\rm turb}$ 
value gives a smaller column density; in the case of 4U 1630$-$47, about 10 
times larger values of $N_{\rm H}$ was obtained with $v_{\rm turb} = 2000$ km
sec$^{-1}$ than that with $v_{\rm turb} = 100$ km sec$^{-1}$ \citep{kub07}. 
If we assumed $v_{\rm turb} \sim 5000$ km sec$^{-1}$, the $R$ values of the
MAXI J1305$-$704 would become much larger than those of GRO J1655$-$40. 
The BHXB GRS~1915$+$105 also shows a large apparent distance
($R \approx 10^6$ km), but its binary size is very large ($\sim 10^8$ km)
as the source has a very long ($33.5 \pm 1.5$ days; \citealt{gre01}) 
orbital period and contains a massive black hole ($\approx 15 M_\sun$). 
Thus, the distance of the ionized absorber in GRS 1915$+$105 would be well 
within the size of accretion disk, even if $R \sim \Delta R$ is assumed.
Figure~\ref{fig_xsabs} plots the values of $N_{\rm H}$ and $R$ for 
the ionized absorbers in MAXI J1305$-$704 and those in other BHXBs 
that exhibit dips and have turbulent velocities less than 1000 km 
sec$^{-1}$, except for GRS 1915$+$105. The absorbers in MAXI 
J1305$-$704 have large $R$ values and comparable or smaller 
$N_{\rm H}$ values compared to those in the other sources.

These properties suggest that the absorbers in MAXI~J1305$-$704 are 
originated from compact structures like those responsible for the dips, rather 
than a typical disk wind, which would be widely extended in the radial
direction (i.e., $\Delta R/R \sim 1$). The {\it Suzaku} light curve in the soft band 
exhibits significant time variability even in the non-dip phases on the shorter 
time scale than the orbital period. We find these small variabilities 
occur almost recurrently as well as the deep and shallow dips, by folding the 
Suzaku light curve with the orbital period. This fact would support the idea that 
the ionized absorbers in MAXI J1305$-$704 are associated with the disk and 
composed of compact clouds, 
unlike disk winds observed in other BHXBs that distribute quite homogeneously
over the orbital phase \citep{yam01}. \citet{mil13} have recently obtained 
a very large number density ($n_{\rm H} \approx 10^{17}$ cm$^{-3}$) from 
density-sensitive absorption lines seen in the high-resolution 
{\it Chandra}/HETG spectrum of the source in the high/soft state and derived   
the actual distance of the absorber as $\approx 4\times 10^3$ km utilizing 
the density without any assumption of $\Delta R/R$. From the distance 
and density, we can estimate the size of ionized absorber as 
$\Delta R \approx 1$ km. This strongly indicates that the absorbers are 
composed of small clumps.  

A possibility is that a fraction of the absorbing gas responsible for the 
dips is spread to the non-dip phases. MAXI~J1305$-$704 is likely to be a 
high inclination system even compared with other dipping BHXBs and we 
may see many complex structures of the absorbing gas with a small scale 
height on the surface of the accretion disk. As described in \citet{fra87}, the 
short orbital period of MAXI J1305$-$704 may maintain the clumpy 
absorbers and produce the similarity in the properties of the ionized absorbers 
seen in neutron star binaries. Also, like neutron star dippers, the disk 
size of MAXI J1305$-$704 ($\lesssim 10^6 M_4^{1/3}$ km), estimated from the 
binary size, is comparable with or maybe smaller than the Compton radius 
($\sim 4 \times 10^5$ $T_{\rm IC8}^{-1}$ $M_4$ km, where 
$T_{\rm IC8}$ is the Compton temperature in units of $10^8$ K) and would not 
be sufficiently large to power a thermal-driven disk wind \citep[see][]{dia12}. However, 
there remains a possibility that we are seeing the launching site of the disk wind 
near the outer edge of the disk, which may be less homogeneous than in 
its outer parts and can be compact ($\Delta R/R < 1$). 
If this were the case, our {\it Suzaku} result would imply that such a disk wind 
exists in the low/hard state of a BHXB and that the accretion states do not 
always determine the presence of disk winds. Future studies using high quality 
and high resolution spectra, like those obtained by ASTRO-H \citep{tak10}, 
should provide important clues to reveal the origin of these ionized absorbers 
in X-ray binaries.

\section{Summary}
The {\it Suzaku}, {\it Swift}, and {\it IRSF} observations of the newly discovered black 
hole candidate MAXI~J1305$-$704 provide us with the following results:

\begin{enumerate}
\item The source clearly shows two absorption dips with different 
mean hardness ratios. 
They have the same interval ($9.74 \pm 0.04$ hours), which likely 
corresponds to the orbital period. 

\item The {\it Suzaku} non-dip spectrum in the low/hard state can be described
with a model composed of a multicolor disk emission, its Comptonization,
and a reflection component, absorbed by an ionized gas.

\item 
The {\it Swift}/XRT spectrum in the high/soft state is well reproduced with a
relativistic disk emission model with a moderate spin parameter
($\approx 0.5$) for an assumed inclination angle of $75^\circ$, a black hole mass 
of 3 $M_\sun$, and a luminosity of $0.05 L_{\rm Edd}$. The inner disk radius obtained
in the {\it Swift}/XRT spectrum is much smaller than that in the {\it Suzaku}
spectrum, indicating that the inner edge of the standard disk was
receded as the state transition from the high/soft state to the low/hard
state.

\item The ionized absorbers in the dip spectra of {\it Suzaku} have smaller
ionization parameters and larger hydrogen column densities than those of
the non-dip spectrum. Similar trends are observed from dipping neutron
star X-ray binaries.

\item 
We find that the ionized absorbers have much smaller ionized
parameters and column densities than those of typical disk winds seen in
other BHXBs. The properties of the absorbing gas are rather similar to
the deeply-dipping neutron star X-ray binaries. These results suggest
that the absorbers have compact and clumpy structures like those
responsible for the dips rather than a homogeneous disk wind. The
possibility that we are seeing the launching site of a disk wind is not
ruled out, however.

\item Near infrared observations in the $J$, $H$, and $K_{\rm s}$ bands
were also performed with {\it IRSF} both in the high/soft state and low/hard state.
The fluxes in the three bands are about an order of magnitude larger than the disk 
emission estimated from the X-ray spectra and can be described with the reprocessed 
thermal emission from the irradiated outer disk and the black body emission from 
the companion star.

\end{enumerate}

\acknowledgments

We thank the {\it Suzaku} operation team for arranging and carrying out the 
TOO observation. This publication made use of data products from the Two
Micron All Sky Survey, which is a joint project of the University of
Massachusetts and the Infrared Processing and Analysis Center/California
Institute of Technology, funded by the National Aeronautics and Space
Administration and the National Science Foundation. This work is partly
supported by a Grant-in-Aid for JSPS Fellows for young researchers (MS)
and for Scientific Research 23540265 (YU) and 19047001 (NK).

{\it Facilities:} \facility{{\it Suzaku}}, \facility{{\it MAXI}}, 
\facility{{\it Swift}}, \facility{{\it IRSF}}.


\clearpage
\setlength{\headsep}{30mm}
\begin{landscape}
\begin{table}
\caption{Properties of ionized absorbers in X-ray binaries \label{tab_wind}}
\scriptsize
\begin{tabular}{lcccccccccccc}
\tableline\tableline
Source name & photon index & distance & orbital period & $N_{\rm H}$ & $\log \xi$ & $L_{\rm X}$ & 
energy band of & $R$ & $v_{\rm outflow}$ & $v_{\rm turb}$ & Reference \\
& or State & (kpc) & (h) & ($10^{22}$ cm$^{-2}$) & &  (erg sec$^{-1}$) &  $L_{\rm X}$ (keV) & 
($10^{5}$ km) & (km s$^{-1}$) & (km s$^{-1}$) & \\
\tableline
XB~1916$-$053 & $2.25 \pm 0.03$ & 9.3 & 0.8 h & $4.2 \pm 0.5$ & $3.05 \pm 0.04$ & $4.4 \times 10^{36}$ & 0.6--10 
& 9.3 & -\tablenotemark{l} & $2300^{+2100}_{-1700}$ & \citet{dia06}\\
4U~1254$-$690 & $2.09 \pm 0.02$ & 10 & 5.8 h &  $8.4 \pm 0.3$ & $4.3 \pm 0.1$ & 
$1.04 \times 10^{37}$ & 0.6--10 & 0.62 & -\tablenotemark{l} & $2800 \pm 1900$ & \citet{dia06}\\
MXB~1659$-$298 & $1.96 \pm 0.03$ & 15 & 7.1 h & $11.1 \pm 0.6$ & $3.8 \pm 0.1$ & 
$3.44 \times 10^{37}$ & 0.6--10 & 4.9 & -\tablenotemark{l} & $700^{+1000}_{-350}$ & \citet{dia06}\\
EXO~0748$-$676 & $1.57 \pm 0.05$ & 10 & 3.8 h & $3.5 \pm 0.2$ & $2.45 \pm 0.02$ & 
$3.4 \times 10^{36}$ & 0.6--10 & 34 & -\tablenotemark{l} & $13 \pm 6$ &\citet{dia06} \\
XB~1323$-$619 & $1.90^{+0.06}_{-0.10}$ & 10 & 2.9 h & $3.6^{+1.0}_{-0.9}$ & $3.90^{+0.08}_{-0.09}$ & 
$5.2 \times 10^{36}$ & 0.6--10 & 1.8 & -\tablenotemark{l} & $1700 \pm 1000$ & \citet{boi05}\\
X~1624$-$490 & 2.25 & $15^{+2.9}_{-2.6}$ & $21$ h & $20 \pm 10$ & $4.3 \pm 0.4$ & $4.9 \times 10^{37}$ 
& 1--10 & 1.2 & $607^{+354}_{-342}$ & $280^{+180}_{-80}$ & \citet{xia09} \\
 & & & &$1.3^{+0.3}_{-0.5}$ & $3.3 \pm 0.2$ & & & 190 & $-213^{+108}_{-158}$ & $<174$ & \\
GX~13$+$1 & MCD dominant & $7 \pm 1$& 24 day\tablenotemark{c} 
& 10--100 & 4.2--4.5 & $1 \times 10^{38}$ & full\tablenotemark{i} & 1--10 &  $460 \pm 70$ & $490^{+110}_{-140}$ & \citet{ued04} \\ 
IGR~J17480$-$2446 & 1.26 & 5.5 & 21 h\tablenotemark{d} & 3\tablenotemark{g} & 3\tablenotemark{g} & $(3.7 \pm 0.2)\times 10^{37}$ & 0.5--10 & 
123 & 3100 & $4800\pm 900$\tablenotemark{m} & \citet{mil11} \\ 
 & & & & 2\tablenotemark{h} & 4.3\tablenotemark{h} & & & 9 & 500 & $<600$\tablenotemark{m} & \\
Cir~X-1 & $0.38^{+0.29}_{-0.19}$ & 6 & 16.6 day & $80 \pm 20$ & $1.6^{+0.4}_{-0.2}$ & $1.4 \times 10^{36}$ & 2--10 
& 4.3 & $2300^{+840}_{-1400}$ & 300 (fixed) & \citet{sch08} \\
  & $2.62 \pm 0.12$ & & & $6 \pm 1$ & $2.7^{+0.1}_{-0.2}$ & $9.9 \times 10^{36}$ 
& 2--10 & 33 & $570^{+670}_{-630}$ & 300 (fixed) & \\
\tableline
4U 1630$-$47 & high/soft state & 10 & unknown & 5--19 & 4.38--4.88 & $2.8 \times 10^{38}$ & 
full\tablenotemark{i} & 0.1--1 & ($-$120)--1740 & 500 (fixed) & \citet{kub07}\\
H 1743$-$32\tablenotemark{a} & high/soft state & 8.5 & unknown & $\approx 5$ & 5.7 & 
$6.8 \times 10^{38}$ & 0.5--10 & 0.01--1 & $670 \pm 170$\tablenotemark{i} & 
$1800 \pm 400$\tablenotemark{l} & \citet{mil06b}\\
GRO J1655$-$40 & high/soft state & 3 & 2.6 day & 30--100 & 3 & $ 1 \times 10^{36}$ & 9--$\infty$ 
& 0.1--0.3 & -\tablenotemark{l} & $<130$\tablenotemark{h} & \citet{ued98}\\
GRO J1655$-$40 & high/soft state & 3.2 & & $5.2 \pm 1.0$ & $3.60 \pm 0.04$ & 
$8 \times 10^{37}$ & 0.5--200 & 40 & $540 \pm 120$ & $3500 \pm 900$& \citet{dia07}\\
 & & & & $1.5\pm 1.2$ & $3.30 \pm 0.04$ &$4 \times 10^{37}$ & & 130& &$5900 \pm 1200$ & \\
GRO J1655$-$40 & high/soft state & 3.2 & & 12--50 & 4.8--5.7 & $5 \times 10^{37}$ & 0.65--10 & 0.1--0.25 
& 300--1600 & 300 (fixed) & \citet{mil08}\\
GRS~1915$+$105& ``low/hard'' state & 12.5 & $33.5 \pm 1.5$ day\tablenotemark{e} & $\approx 1$ & $\approx 3.8$ 
& $4 \times 10^{38}$ & 2--10 & $\approx 10$ & $\approx 1000$ & 740\tablenotemark{h} (fixed) & \citet{kot00} \\
GRS~1915$+$105& ``low/hard'' state & 12.5 & & $\approx 3$ & $\lesssim 4.15$ & 
$6.4 \times 10^{38}$ & full\tablenotemark{i} & $\lesssim 20$\tablenotemark{j} & $700\pm 400$\tablenotemark{i} 
& $578\pm 400$\tablenotemark{h} & \citet{lee02} \\
GRS~1915$+$105 & high/soft state & 12 & & $\approx 10$ & 4.2--4.3 &
 (6.6--8.3) $\times 10^{38}$ & 0.01--100 & 2--6 & 90--560 & 70--200 & \citet{ued09} \\
GX 339$-$4 & intermediate state\tablenotemark{b} & 5 & 1.76 day\tablenotemark{f} & 0.02 & $\approx$ 3 & 
$\approx 1 \times 10^{37}$ & 0.5--10 & 20\tablenotemark{k} & 510\tablenotemark{n} & 410\tablenotemark{n} & \citet{mil04} \\
MAXI J1305$-$704 & high/soft state & 6 & $9.74\pm 0.04$ h& $5.1^{+3.8}_{-2.8}$ & $2.86^{+0.52}_{-0.18}$ 
& $8.1 \times 10^{36}$ & 0.0136--13.6 & $22$ & $< 4800$ & 300 (fixed) & This work\\ 
MAXI J1305$-$704 & high/soft state & & & $1.0^{+0.4}_{-0.3}$ & $1.2 \pm 0.2$ & & & 
$5.1 \times 10^3$ & 0 (fixed) & 300 (fixed) & \\
MAXI J1305$-$704 & low/hard state & & & & & & & & & & \\
 & (non-dip) & & & $0.61^{+0.09}_{-0.10}$ & $2.18 \pm 0.04 $ & $1.3 \times 10^{36}$ & & $140$ & $<2300$ & 300 (fixed) & This work\\
MAXI J1305$-$704 & (deep dip) & & & $14.4 \pm 0.6$ & $1.90 \pm 0.07 $ & & & $11$ & $<2700$ & 300 (fixed) & \\
MAXI J1305$-$704 & (shallow dip) & & & $6.6^{+0.5}_{-0.4}$ & $1.79 \pm 0.07 $ & & & $32$ & $<5800$ & 300 (fixed) & \\
\tableline
\end{tabular}
\tablenotetext{a}{Observation 1 in \citet{mil06b}.}
\tablenotetext{b}{In the soft-to-hard transition.}
\tablenotetext{c}{\citet{cor10}.}
\tablenotetext{d}{\citet{pap11}.}
\tablenotetext{e}{\citet{gre01}.}
\tablenotetext{f}{\citet{hyn03}.}
\tablenotetext{g}{Estimated from the Fe XXV absorption line.}
\tablenotetext{h}{Estimated from the Fe XXVI absorption line.}
\tablenotetext{i}{Bolometric luminosity.}
\tablenotetext{j}{$\Delta R/R \approx 0.1$.}
\tablenotetext{k}{Calculated by assuming the thickness and number density 
of the absorber as $20$ km and $8 \times 10^{13}$ cm$^{-3}$, respectively.}
\tablenotetext{l}{Not constrained.}
\tablenotetext{m}{The line width when the Gaussian model is applied.}
\tablenotetext{n}{The results of the Ne II line at $14.631$ \AA, from which the largest blue shift is obtained.}

\tablecomments{$R$ is calculated from $L_{\rm X}/\xi N_{\rm H}$ by assuming $\Delta R = R$, unless 
otherwise stated. The positive values of $v_{\rm outflow}$ indicate blue shifts.
The distances represent the assumed values that are used to estimate $L_{\rm X}$. 
4U 1630$-$47 and the following sources are BHXBs. The results in the non-dip phases are shown 
for the top 6 sources (dippers).}
\end{table}
\clearpage
\end{landscape}

\end{document}